# Discovery of unidirectional charge and pair-density-wave orders in topological monolayer 1T′-MoTe$_2$

Li-Xuan Wei[1,†], Peng-Cheng Xiao[2,†], Fangsen Li[3,†], Li Wang[3], Bo-Yuan Deng[3], Fang-Jun Cheng[1], Fa-Wei Zheng[4], Ning Hao[5], Ping Zhang[2,6,*], Xu-Cun Ma[1,7,*], Qi-Kun Xue[1,7,8,9], Can-Li Song[1,7,*]

[1]*State Key Laboratory of Low-Dimensional Quantum Physics, Department of Physics, Tsinghua University, Beijing 100084, China*

[2]*School of Physics and Physical Engineering, Qufu Normal University, Qufu 273165, China*

[3]*Vacuum Interconnected Nanotech Workstation, Suzhou Institute of Nano-Tech and Nano-Bionics, Chinese Academy of Sciences, Suzhou 215123, China*

[4]*Key Lab of Advanced Optoelectronic Quantum Architecture and Measurement, School of Physics, Beijing Institute of Technology Beijing 100081, China*

[5]*Anhui Province Key Laboratory of Condensed Matter Physics at Extreme Conditions, High Magnetic Field Laboratory, HFIPS, Chinese Academy of Sciences, Hefei 230031, China*

[6]*Institute of Applied Physics and Computational Mathematics, Beijing 100088, China*

[7]*Frontier Science Center for Quantum Information, Beijing 100084, China*

[8]*Shenzhen Institute for Quantum Science and Engineering and Department of Physics, Southern University of Science and Technology, Shenzhen 518055, China*

[9]*Beijing Academy of Quantum Information Sciences, Beijing 100193, China*

Symmetry-breaking phenomena are observed in numerous strongly-correlated systems including high-temperature superconductors. However, identifying these exotic phases and understanding their interplay with superconductivity in topological materials remain challenging. Here we employ a cryogenic scanning tunneling microscopy to discover a unidirectional (striped) charge order (CO) and a primary pair-density-wave (PDW) in topological monolayer 1T′-MoTe$_2$. The two orders are spatially modulated uniaxially and share the same periodicity of five-unit cells or a wavevector $Q_C \approx Q_P \approx (0, 0.4)\pi/b$ ($b$ is the lattice constant along the Mo chains). Importantly, the primary PDW state features a two-gap superconductivity below the transition temperature of 6.0 K and induces another unique particle-hole-symmetric CO at twice the PDW wavevector. Combining these results and our density functional calculations, we reveal that the two striped orders are primarily driven by nesting behaviors between electron and hole pockets. Our findings establish monolayer 1T′-MoTe$_2$ as a topological paradigm for exploring multiple preexisting symmetry-breaking states.

[†]These authors contributed equally: Li-Xuan Wei, Peng-Cheng Xiao, Fangsen Li

[*]Correspondence to: zhang_ping@iapcm.ac.cn, xucunma@mail.tsinghua.edu.cn, clsong07@mail.tsinghua.edu.cn

**Introduction**

Layered transition metal dichalcogenides $MX_2$ (M = Mo, W; X = S, Se, Te) in the distorted octahedral structure (1T′ and $T_d$) have attracted significant research interest as atomic-scale building blocks for various electronic phenomena, including the quantum spin Hall (QSH) state[1-5], topological Weyl semimetal state[6], superconductivity[7-11] and ferroelectricity[12,13]. These materials connately merge electron correlation and topology, two prominent issues in condensed matter physics that were traditionally explored independently. This integration holds promise for a multitude of quantum states that go beyond the individual components. For example, there has been a particular attention on combining the superconductivity and QSH state to pursue Majorana zero modes (MZM)[14,15], with potential applications for fault-tolerant quantum computation. Despite the growing interest in this area, the charge order (CO) instabilities and the pair-density-wave (PDW) state with modulated superconducting order parameter $\Delta(r)$ ($r$ represents a position), which were recently explored intensively in correlated electronic systems[16-24], has been very limitedly investigated in $MX_2$[25-27]. Consequently, a comprehensive understanding of the CO instability and its relationship with other symmetry-breaking states, particularly the superconductivity, in $MX_2$ with topological nontrivial properties remains to be elucidated.

We achieve these objectives by utilizing the state-of-the-art molecular beam epitaxy (MBE) to prepare monolayers (ML) of metastable 1T′-$MoTe_2$ on graphitized SiC(0001) substrates[28]. Through optimization of the growth conditions, we succeed in achieving single-phase ML 1T′-$MoTe_2$ films with high homogeneity and lateral dimensions on the scale of tens of nanometers (see Methods and Supplementary Fig. 1 for more details). While doping ML 1T′-$MoTe_2$ with electrons was proposed to induce superconductivity followed by a nonsymmorphic 2 × 5 CO[25], it is noteworthy that $T_d$-$MoTe_2$ itself is superconducting below a transition temperature $T_c$ ~ 0.1 K in bulk[7]. This $T_c$ can be enhanced under increased pressure[7] or by chemical substitution[8]. Further studies have revealed signatures of two-gap superconductivity and a reduction of $T_c$ by disorder[8,10], suggesting an unconventional superconductivity in $MoTe_2$. It has been recently found that the in-plane critical field exceeds the Pauli paramagnetic limit and possesses a two-fold symmetry in few-layer $T_d$-$MoTe_2$[11]. As the bulk 1T′($T_d$)-$MoTe_2$ is thinned down to the ML limit[9], there is a remarkable 60-fold enhancement in $T_c$, implying the essential role of electronic correlations in driving the enhancement of $T_c$. In this work, we focus on the topological MLs of 1T′-$MoTe_2$. Using spectroscopic imaging scanning tunneling microscopy (SI-STM), we visualize an unprecedented CO and a primary PDW order, which are both spatially modulated in a unidirectional manner.

**Results**

**Atomic and electronic structures of ML 1T′-$MoTe_2$.** In 1T′-$MoTe_2$, Mo atoms are distorted from the perfect octahedral sites, forming one-dimensional (1D) zigzag chains along the $b$ axis and doubling the period along the orthogonal $a$ axis (see inset of Fig. 1a). Such a distortion lowers the energy levels of Mo-$d$ orbitals and induces a band inversion with Te-$p$ orbitals. Consequently, ML 1T′-$MoTe_2$ has topological nontrivial behaviors[1], consistent with our calculations based on density functional theory (DFT) in Fig. 1a. Owing to relatively weak spin-orbit coupling, the ML 1T′-MoTe$_2$ is a semimetal with a negative band gap of ~ 0.21 eV. The Fermi surface (FS) consists of a hole pocket centered at Γ and two electron pockets on either side of Γ along the reciprocal Γ-Y ($q_y$) axis (Fig. 1b), matching nicely with previous band structure measurements via angle-resolved photoemission spectroscopy[29].

Figure 1c represents a STM topography $T(r)$ with atomic resolution, revealing the distinctive zig-zag chain structure. Using the lattice constant of the uncovered graphene (2.46 Å) as a reference, we determine the average in-plane lattice constants of ML 1T′-MoTe$_2$ as $a = 6.5 \pm 0.1$ Å and $b = 3.4 \pm 0.1$ Å by analyzing STM topographies from various samples and regions. The differential conductance spectra $g(r, V) \equiv dI/dV(r, V)$, which probe the quasiparticle density of states (DOS) as a function of energy $E = eV$ at a given position $r$ ($e$ denotes the elementary charge), display spatial homogeneity across the ML 1T′-MoTe$_2$ films (Fig. 1d, Supplementary Fig. 1b). They are characterized by finite spectral weights at the Fermi level ($E_F$), as expected for a semimetal. A prominent DOS peak exists near -0.65 eV and aligns in energy with onsets of two parabolic hole bands in ML 1T′-MoTe$_2$[29]. More remarkably, we combine our SI-STM technique and DFT calculations to reveal the present QSH states at step edges (Fig. 1e,f), as detailed in Supplementary Fig. 2-5 and Note 1. They are located predominantly below $E_F$ (Fig. 1f) and spread over a lateral dimension of ~ 2.8 nm (Fig. 1e), in line with previous reports in ML 1T′-MoTe$_2$[29] and its sister compounds such as 1T′-W(Te, Se)$_2$[2,4,5].

**Direct visualization of a unidirectional CO.** In addition to the QSH states, we find spatial stripe-like modulations that run unidirectionally along the $b$ axis in ML 1T′-MoTe$_2$. This can be directly visualized in a large-scale $T(r)$ (Fig. 2a) and confirmed by its Fourier transform amplitude $T(q)$ (where $q$ is the wavevector) (Fig. 2b). The stripes have a spacing of $17.1 \pm 0.4$ Å that closely matches five ($\pm$ 0.2) times the lattice constant $b$ (~ $5b$). This nearly commensurate and long-range striped pattern remains largely unaffected in periodicity by twist angles between the 1T′-MoTe$_2$ and underlying graphene (Supplementary Fig. 6), which excludes the stripes originating from any Moiré fringes. Instead, our results indicate emergence of a striped CO at wavevector $Q_C \approx (0, 0.4)\pi/b$.

In the CO phase, the DOS is expected to undergo rearrangement, forming alternating stripes of charge accumulation and depletion in real space. This results in enhanced (suppressed) intensities in the filled-state $dI/dV$ maps $g(r, V < 0)$ over the charge accumulation (depletion) stripes, and vice versa in the empty-state ones $g(r, V > 0)$. We show such a contrast inversion of $g(r, V)$ between opposite biases in Fig. 2c and more details in Supplementary Fig. 7, hallmarks of a CO. To further support this claim, we analyze spectroscopic imaging data of the $V$-dependent $g(r, V)$ in a 33 nm × 33 nm FOV. Linecuts of $g(q, V)$, the amplitude of Fourier transform of $g(r, V)$, along the $q_y$ axis with varying $V$ from -0.3 eV to 0.3 eV reveal that the $Q_C$ is nondispersive with $E = eV$ and fixed at $Q_C \approx (0, 0.4)\pi/b$ (Fig. 2d). This finding strongly indicates that the unidirectional DOS modulation is derived from an exotic CO, rather than quasiparticle interference where the wavevector $q$ of electron scatterings alters with $E$. This CO opens a partial energy gap $\Delta_C \sim 0.14$ eV near $E_F$ and gets suppressed at an elevated temperature (Supplementary Fig. 8). We note the most pronounced contrast inversion of $g(r, \pm V)$ occurring within $\Delta_C$ (i.e., $|V| \leq \Delta_C/2e$, Supplementary Fig. 7d), as expected when attributing the stripes to CO.

Our direct visualization of the unidirectional CO contrasts sharply with recent predictions of electron doping (light)-tunable $2 \times 5$ ($2 \times 2$) COs in 1T′-MoTe$_2$[25,27]. To understand its driving mechanism, we calculate the Lindhard response function, which reduces into the bare electronic susceptibility $\chi(q) = \chi'(q) + i\chi''(q)$ without the renormalization effect. $\chi'(q)$ and $\chi''(q)$ between the electron and hole bands are presented in Fig. 2e,f, and detailed in Supplementary Fig. 9. We neglect the intra-band components due to their negligible strength in comparison with the interband ones. At a low but wide electron doping $n_e$, i.e., $n_e = 0.05e$ per unit cell ($\sim 2.3 \times 10^{13}$/cm$^2$), both real part ($\chi'(q)$) and imaginary part ($\chi''(q)$) of the $\chi(q)$ exhibit the strongest peaks along the $q_y$ axis near $\pm 0.4\pi/b$ (i.e., $Q_C$). This renders us conclude that the Lindhard response function satisfying interband nesting conditions is implicated in driving the CO at $Q_C$. Moreover, the robustness of the CO implies a finite but insignificant electron transfer between the van der Waals stacked graphene and ML 1T′-MoTe$_2$.

**Two-gap superconductivity**. We proceed to study the low-energy d$I$/d$V(r, V)$ spectra at 0.4 K, well below $T_c$, which reveal two pairs of superconducting energy gaps with particle-hole symmetric coherence peaks (Fig. 3a). The vertical dashes indicate the positions of the smaller and larger gaps, measured statistically, at $\Delta_1 = 1.27 \pm 0.08$ meV and $\Delta_2 = 1.91 \pm 0.16$ meV, respectively. Although the coherence peaks corresponding to $\Delta_1$ are observable in all spectra measured, the gap edges of $\Delta_2$ occasionally appear blurred and become hardly distinguishable (Supplementary Fig. 10). This contrast might correlate with band-selective tunneling matrix elements[30]. We hereafter focus on $\Delta_1$ that gets smeared out at elevated temperatures and almost disappears at $T_c \sim 6.0$ K (Fig. 3b). We evaluate the reduced gap $2\Delta_1/k_B T_c \sim 4.9$ ($k_B$, the Boltzmann constant), indicating a strong-coupling superconductivity. This is strongly supported by our DFT calculations, which reveal significantly broadened linewidths of the phonon spectra in ML 1T′-MoTe$_2$ compared to those of the bulk counterpart (Supplementary Fig. 11).

It is worth noting that all the superconducting gaps exhibit rounded bottom near $E_F$, which fail to be straightforwardly described by the Dynes model with either nodeless $s$-wave or $d$-wave paring function[31]. Meanwhile, the deviation cannot be simply attributed to disorder effects, because they, if present, would have significantly attenuated the coherence peaks (see Supplementary Note 2 and Fig. 12). Alternatively, we emphasize that the finite spectral weights within $\Delta_1$, especially close to $E_F$, are intrinsic to ML 1T′-MoTe$_2$ and match a partially gapped FS in the superconducting state. Upon application of an out-of-plane magnetic field $B$, the ingap spectral weights are gradually increased until a pseudogap-like feature remains above an upper critical field $B_{c2} \sim 4$ T (Fig. 3c). Thermal broadening of the spectrum at 5 T to a temperature of 7.0 K yields a curved background (see the red dashes in Fig. 3c), which reproduces the measured d$I$/d$V$ spectrum at 7.0 K and zero field in Fig. 3b. Using the Ginzburg-Landau formula $B_{c2} = \Phi_0/2\pi\xi^2$ ($\Phi_0$, the flux quantum), we estimate the coherence length $\xi \sim 9.1$ nm.

**Striped PDW order**. Most strikingly, we discover a remarkable spatial modulation of the $\Delta_1(r)$ in a unidirectional manner, absent in ordinary (uniform) superconductors. This corresponds to the emergence of a striped PDW order. As illustrated in Fig. 3d,e, a distinct variation exists in $\Delta_1(r)$ across the alternating charge accumulation and depletion stripes, which are spaced only 8.6 Å (~ 2.5$b$) apart in the unidirectional CO state. The superconducting gaps exhibit larger $\Delta_1(r)$ on the charge depletion stripes, accompanied by attenuated coherence peaks and ingap DOS, and vice versa on the charge accumulation stripes. To see this more explicitly, we collect a series of d$I$/d$V$ spectra along a trajectory marked by the white arrow in Fig. 3d. Evidently, $\Delta_1(r)$ displays pronounced spatial modulations with the same 5$b$ periodicity as the CO (Fig. 3f), characterized by $Q_P \approx (0, 0.4)\pi/b$. Conversely, no periodic modulation is observable in $\Delta_1(r)$ along the crystal $a$ axis (Supplementary Fig. 13). The striped PDW features Cooper pairs with nonzero center-of-mass momentum $P = \hbar Q_P$ ($\hbar$ is Planck's constant $h$ divided by $2\pi$) and differs from uniform Bardeen-Cooper-Schrieffer (BCS) superconductors. It is sensitive to quenched disorders (Supplementary Note 2) and typically has ungapped FS portions (Fig. 3a-c, Supplementary Fig. 10, 12 and 13)[32,33], as exhibited by finite ingap spectral weights near $E_F$ at 0.4 K. Such a striped PDW has been long proposed in the canonical cuprate superconductor La$_{2-x}$Ba$_x$CuO$_4$[34] as a primary explanation for broad experimental anomalies in cuprates, but not yet unambiguously visualized so far[18,19]. Recently, the striped PDW order was suggested at domain walls of ML Fe(Te, Se) films[21] and in magnetic superconductor EuRbFe$_4$As$_4$[22], leaving their microscopic origin unknown.

To investigate the striped PDW quantitatively and its possible relationship with the preexisting CO, we extract the magnitude $\Delta_1(r)$ of the superconducting gaps from the coherence peak positions (see details in Supplementary Fig. 14), the coherence peak height $H(r)$ as averaged $g(r, V = \pm\Delta_1/e)$ and integrated DOS $I(r, |V| \leq \Delta_1/e)$ within $\Delta_1(r)$. These quantities characterize the strength of Cooper pairs, superconducting phase coherence, and unpaired electrons within $\Delta_1(r)$ due to the ungapped FS. They are fundamentally significant for describing the PDW state. Figure 3g presents these parameters $\Delta_1(r)$, $H(r)$ and $I(r)$, along with the CO modulation $g(r, 0.3$ eV$)$, obtained from the spatially-resolved d$I$/d$V(r, V)$ data in Fig. 3d-f. All of them oscillate in space with the same periodicity of ~ 5$b$, which we have further confirmed by mapping the PDW order in Supplementary Fig. 15. Most significantly, all raw $g(r, V)$ maps with $|V| < \Delta_1/e$ exhibit spatial modulations at $Q_P$, even for $V = 0$ mV, with a particle-hole symmetry (Supplementary Fig. 16). The dichotomy between the PDW (particle-hole symmetry) and the CO (particle-hole asymmetry) excludes the cause of the spatially-modulated $g(r, V)$ within $\Delta_1$ by the CO. Instead, they reflect a coupling of the striped PDW order $\Delta_Q$, namely the gap modulation amplitude at $Q_P$, to a uniform superconducting order parameter $\Delta_0$ (i.e. $\Delta_Q\Delta_0^*$). This leads to spatial distributions of the $g(r, V)$ maps in the primary PDW state, as what the $I(r)$ describes, which are out of phase with the $\Delta_1(r)$ (Fig. 3g and Supplementary Fig. 15).

**Primacy of PDW and its induced CO at $2Q_P$.** The significant modulation of $\Delta_1(r)$ and $g(r, V \sim 0 \text{ mV})$ on a length scale of $5b$, five times smaller than $\xi \sim 9.1$ nm, provides compelling evidence against the striped PDW being a subsidiary order of the CO state, caused by the coupling of the unidirectional CO to the $\Delta_0$. In this case, the gap modulation is generally negligible[35], e.g., $< 0.01\Delta_0$ in NbSe$_2$[36], and both disorder sensitivity and spatial modulations of $g(r, V \sim 0)$ would be canceled when the FS is fully gapped by $\Delta_0$. Alternatively, our observations are consistent with a primary striped PDW in ML 1T′-MoTe$_2$, i.e., $\Delta_1(r) = F_P\Delta_Q(e^{iQ_P \cdot r} + e^{-iQ_P \cdot r})$, where $F_P$ is the form factor of $\Delta_Q$. Under this context, the Ginzburg-Landau theory for coupled order parameters as product of $\Delta_Q$ with itself $\Delta_Q\Delta^*_{-Q}$ predicts an induced CO modulation at twice the PDW wavevector $Q_P$[18,19,34], i.e., $2Q_P$, which is indeed observed in Fig. 2d. As marked by the green dashes, the $2Q_P$ CO is exclusively discernible near $E_F$ and intimately links to the striped PDW order at $Q_P$, rather than the second harmonic of $Q_C$ (see Supplementary Note 3 and Fig. 17).

To better understand the induced CO at $2Q_P$, we perform Gaussian filters of the low-energy $g(q, E = eV)$ by keeping only the Fourier components around $\pm Q_C$ or $\pm 2Q_P$ with a cut-off length of 0.06 Å$^{-1}$ and then the inverse Fourier transform (IFT). This cut-off length is greater than three times of the widths of both $\pm Q_C$ ($\sim 0.019$ Å$^{-1}$) and $\pm 2Q_P$ ($\sim 0.015$ Å$^{-1}$) peaks and covers most $q$ fluctuations near $\pm Q_C$ or $\pm 2Q_P$. This allows us to distinguish the respective spatial modulations of the preexisting and induced COs, dubbed as $g(r, Q_C, V)$ and $g(r, 2Q_P, V)$, respectively. The $g(q, \pm 10 \text{ mV})$ maps and the resultant IFT images extracted from the raw $g(r, \pm 10 \text{ mV})$ maps (Supplementary Fig. 18) are displayed in Fig. 4a-d. As anticipated, the spatial modulations of $Q_C$ are reversed in intensity between $g(r, Q_C, 10 \text{ mV})$ and $g(r, Q_C, -10 \text{ mV})$. By contrast, the IFT images of $g(r, 2Q_P, \pm 10 \text{ mV})$ present no contrast inversion as the bias polarity is switched, as the striped PDW order does (Supplementary Fig. 16). Such a particle-hole dichotomy between the $Q_C$ and $2Q_P$ CO is more apparently seen in Fig. 4e, which plots the space ($r$)-resolved intensity modulations of all IFT images along an identical trajectory indicated in Fig. 4c. The findings provide conclusive evidence for the primacy of the discovered PDW in ML 1T′-MoTe$_2$. It induces a composite CO at $2Q_P$ that is clearly distinguishable from the preexisting CO at $Q_C$. Given that the $Q_P$ approximately connects the edges of the electron and hole pockets along the $k_y$ axis (Fig. 1b), our results indicate that the interband electron pairing with a finite center-of-mass momentum $\hbar Q_P$ is behind the striped PDW, which has been visualized as unidirectional spatial modulations of $\Delta_1(r)$ in real space.

**Discussion**

Overall, our atomic-scale STM measurements, in conjunction with DFT calculations, have revealed the coexistence of a fascinating CO and a primary PDW in a non-magnetic 2D topological insulator ML 1T′-MoTe$_2$. Based on model calculations[37], axion quasiparticles can emerge from topological materials with CO instabilities. From this point of view, our experimental observation of the CO in ML 1T′-MoTe$_2$, and by implication in the bulk counterpart T$_d$-MoTe$_2$ with topological Weyl semimetal states[6], opens new opportunities for fundamental studies and potential applications of axion electrodynamics. However, it should be emphasized that the underlying physics of ML 1T′-MoTe$_2$ goes beyond its potential as an axion insulator. The CO might potentially modify the topological band structure in a uniaxial way, which warrants further investigation. Moreover, the unidirectional CO opens an energy gap around $E_F$ (Supplementary Fig. 8), which, together with the pseudogap-like feature (Fig. 3b,c), suppresses the bulk electronic DOS. This renews hope for experimental explorations and potential applications of the QSH effects in semi-metallic 1T′-MoTe$_2$ where the QSH edge channels are typically mixed into the bulk states[29].

More importantly, our direct visualization of the striped PDW provides a straightforward explanation for the observed twofold symmetry of in-plane critical field in few-layer MoTe$_2$[11], while the retained superconductivity at the topological step edges (see Supplementary Note 4 and Fig. 19) realizes a potential connate 1D topological superconductor ML 1T′-MoTe$_2$, without the need of constructing complex heterostructures[15]. To corroborate this state, one can couple the edges of ML 1T′-MoTe$_2$ to ferromagnetic clusters. This breaks the time-reversal symmetry[38] and may localize MZMs that could be easily identified via STM as zero-bias conductance peaks within $\Delta_1$. Moreover, the intricate intertwining between the striped PDW and CO induces a unique gap-coherence relationship reminiscent of cuprate and iron pnictides, where electronic nematicity or smecticity was often intertwined with the high-$T_c$ superconductivity. This analogy may originate from the susceptibility of these low carrier density superconductors to the spatial fluctuations between the BCS (charge accumulation stripes with strong phase coherence) and BEC (charge depletion stripes with attenuated phase coherence but enlarged gap) limits[9,39]. Partial melting of the striped PDW order, regardless of whether thermal or quantum, can induce a potential composite charge-4$e$ uniform superconducting order $\Delta_{4e}(r) \propto \Delta_Q \Delta_{-Q}$[34,40]. Though these unusual states ask for further experimental studies, ML 1T′-MoTe$_2$ has already shown itself to be a unique platform that exhibits a wide range of quantum states resembling those revealed in cuprate and iron-based superconductors. It thus holds great promise for finding solutions to the long-standing mysteries in high-$T_c$ superconductors, including electronic nematicity, CO, PDW, intertwined electronic orders, pseudogap phenomenology, nontrivial topology, and the potential impacts on pairing symmetry and unconventional superconductivity.

**Methods**

**Sample growth.** Nitrogen-doped 4$H$-SiC(0001) wafers with a resistivity of 0.015 Ω·cm ~ 0.028 Ω·cm were graphitized by heating them to 1400°C for 10 min in ultra-high vacuum to achieve layer-varying graphene substrates from single-layer to bilayer and multiplayers[41]. High-purity Mo (99.95%) and Te (99.9999%) sources evaporated from a single-pocket electron beam evaporator and standard Knudsen diffusion cell, respectively, were then co-deposited on the graphene/SiC substrates in an extremely high Te/Mo flux ratio (>20). This compensates the loss of volatile Te molecules during the growth, while the excess Te molecules cannot be incorporated into the films and readily desorb at a moderate substrate temperature of 300°C. Meanwhile, this substrate temperature was chosen to mostly depress the predominant growth of the more stable 2*H*-MoTe$_2$ phase and thus allow for the controlled growth of large-sized ML of 1T′-MoTe$_2$ films. A lower substrate temperature renders the ML 1T′-MoTe$_2$ films less integral and decorated by more sheets of bilayer MoTe$_2$. To further enhance the surface cleanliness for STM measurements, we annealed the as-prepared samples under the Te atmosphere at 300ºC for 20 minutes before transferring into STM head.

*In-situ* **STM measurement.** Our STM measurements were carried out in a commercial Unisoku USM 1300 system interconnected to an MBE chamber for *in-situ* sample preparation. The base pressure for both chambers is lower than $2.0 \times 10^{-10}$ Torr. A maximum magnetic field of 8 T (2 T) perpendicular (parallel) to the sample surface can be applied if needed. Polycrystalline PtIr tips were conditioned by *e*-beam bombardment in UHV, calibrated on the MBE-prepared Ag/Si(111) films, and used throughout the experiments. All STM topographies $T(r)$ were measured in a constant current mode, while the tunneling $dI/dV$ spectra and conductance maps $g(r, V)$ were taken using a standard lock-in technique with a small *a.c.* modulation voltage at a frequency of $f = 983$ Hz.

**DFT calculation.** The first-principles calculations were carried out by using density functional theory (DFT) and density functional perturbation theory (DFPT)[42,43]. We used QUANTUM ESPRESSO (QE) package[44] for electronic structure and lattice dynamics calculations, including phonon spectra and electron-phonon coupling. The McMillan-Allen-Dynes formula[45] in the QE package[44] was used to calculate the physical quantities related to superconductivity. The iterative Green's function was used to compute the surface states implemented in the WANNIERTOOLS package[46], which uses the maximally localized Wannier function (MLWF)[47] based on the VASP2WANNIER90 interfaces and the tight binding method[48].

**Data availability**

Data that support the finding of this study are available from the corresponding authors upon reasonable request. Source data are provided with the paper.

**Acknowledgments**

We thank J. Wang and H. Yao for helpful discussions. The work is financially supported by grants from the National Key Research and Development Program of China (Grant No. 2022YFA1403100), the Natural Science Foundation of China (Grant No. 12141403, Grant No. 52388201 and Grant No. 12204512), the Innovation Program for Quantum Science and Technology (grant No. 2021ZD0302502), and Nano-X from the Suzhou Institute of Nano-Tech and Nano-Bionics (SINANO), the Chinese Academy of Sciences.

**Author contributions**

C.L.S., X.C.M., Q.K.X. and L.X.W. conceived and designed the experiments. L.X.W., L.W., F.S.L. and B.Y.D. carried out the MBE growth and STM measurements. L.X.W. and C.L.S. analyzed the experimental data and plotted the figures. P.C.X., F.W.Z., N. H. and P. Z. carried out the theoretical calculations. C. L. Song., L. X. W., P. C. X. and N. H. wrote the manuscript with comments from all authors.

**Competing interests**

The authors declare no competing interests.

**Additional information**

Supplementary information is available for this paper.

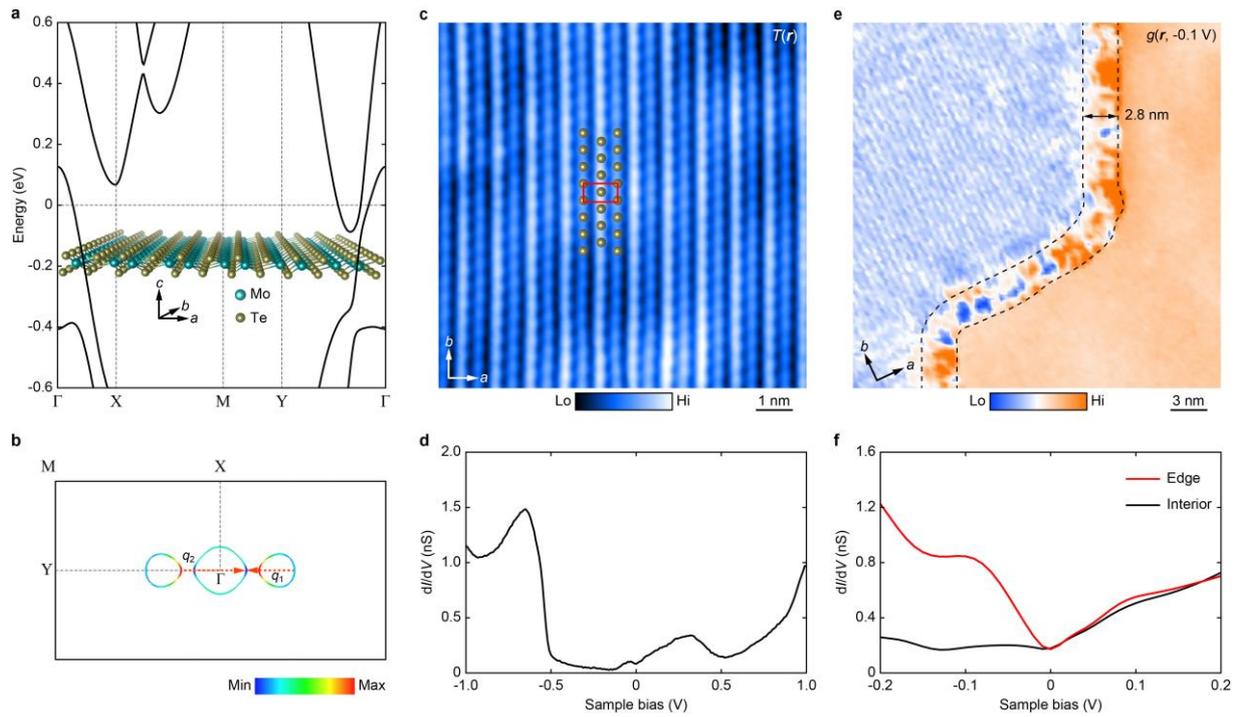

**Figure 1 | Atomic and band structures of ML 1T′-MoTe$_2$. a** Calculated band structure of ML 1T′-MoTe$_2$ with spin-orbit coupling included. Inset shows the schematic crystal structure of ML 1T′-MoTe$_2$. **b** FS of ML 1T′-MoTe$_2$ showing two interband nesting $q_1$ and $q_2$ marked by red arrows. The color gradient measures the magnitude of the Fermi velocity, while the black rectangle denotes the Brillouin zone (BZ) throughout. **c** Atomically-resolved STM topography $T(r)$ (7 nm × 7 nm, $V$ = 5 mV, $I$ = 1 nA) of the 1T′-MoTe$_2$ surface, partially overlaid by the top-layer Te atoms (olive balls). The $a$ and $b$ crystal axes are perpendicular and parallel to the zigzag chains, respectively. The red rectangle encloses a surface unit cell. **d** Representative d$I$/d$V$ spectrum showing a semi-metallic behavior in ML 1T′-MoTe$_2$. Set point: $V$ = 1.0 V, $I$ = 0.5 nA. **e** Conductance map $g(r, V$ = -0.1 V) in a field of view (FOV) of 30 nm × 30 nm FOV, with coexisting ML 1T′-MoTe$_2$ (top-left) and graphene surface (bottom-right). The QSH states emerge along the step edges that are surrounded by the black dashes. Setpoint: $V$ = 0.2 V, $I$ = 0.2 nA. **f** Spatially averaged $dI/dV$ spectra measured at the interiors (black) and step edges (red) of ML 1T′-MoTe$_2$.

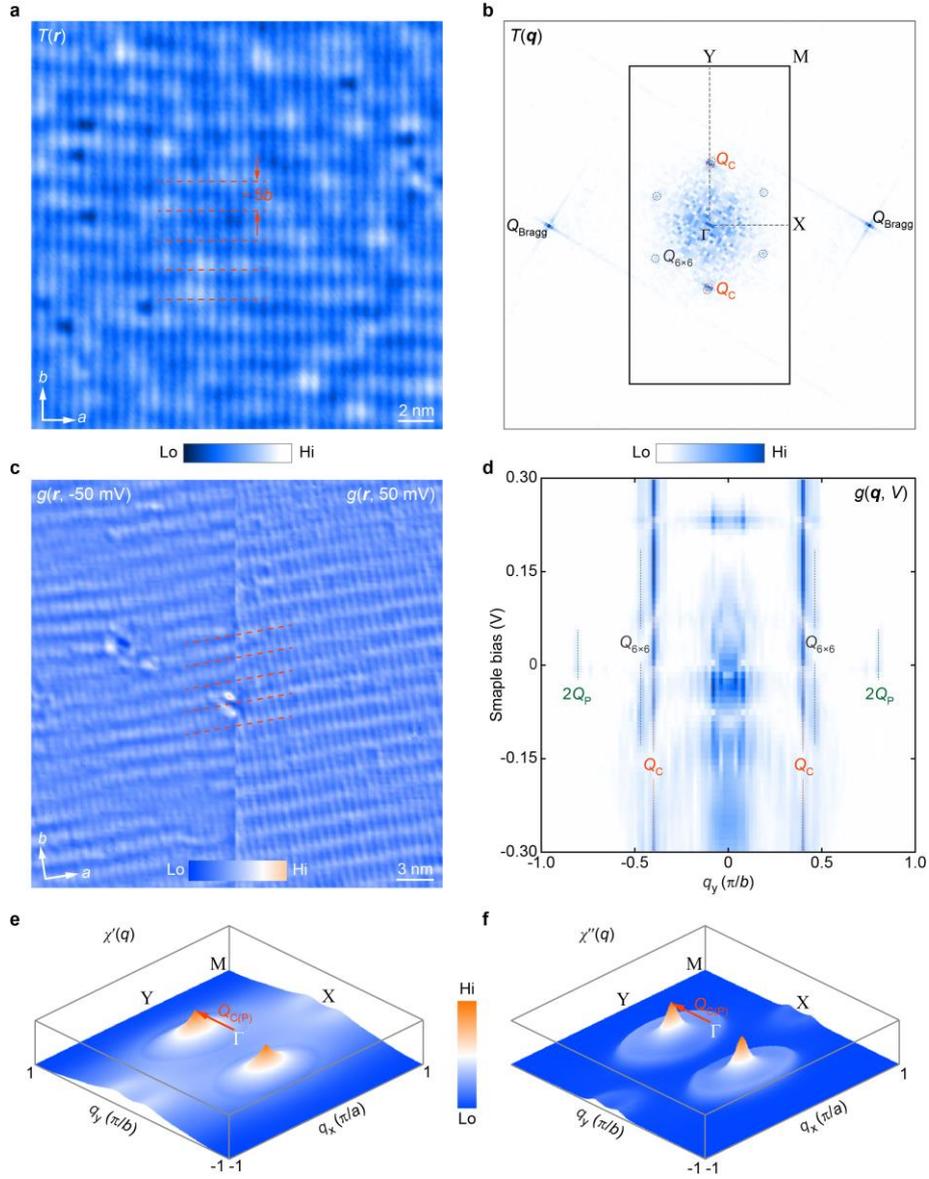

**Figure 2 | Emergence of unidirectional CO. a** Large-scale $T(r)$ (23 nm × 23 nm, $V$ = 1.0 V, $I$ = 10 pA) presenting a unidirectional CO (marked by the red dashes) along the $b$ axis at five-unit-cell periodicity (~ 5$b$) in ML 1T′-MoTe$_2$. **b** Amplitude $T(q)$ of the Fourier transform of $T(r)$ in **a**, revealing wavevectors $Q_C \approx (0, 0.4)\pi/b$ from the unidirectional CO and $Q_{6\times6}$ from the buried substrate. **c** Filled-state (left) and empty-state (right) $g(r, V = \pm50$ mV$)$ maps, showing a contrast inversion of the CO between opposite biases. **d** Bias-dependent amplitude profiles (along the $q_y$ direction) of the Fourier transform of $g(r, V)$, measured on a 33 nm × 33 nm FOV. Vertical dashes denote $Q_{6\times6}$, the wavevectors $Q_C$ of preexisting CO and $2Q_P$ induced by a primary PDW order. **e,f** Real and imaginary electron susceptibility between the electron and hole pockets crossing the $E_F$ with $n_e$ = 0.05$e$ per unit cell, where $Q_C = Q_P$ is indicated by the red arrows.

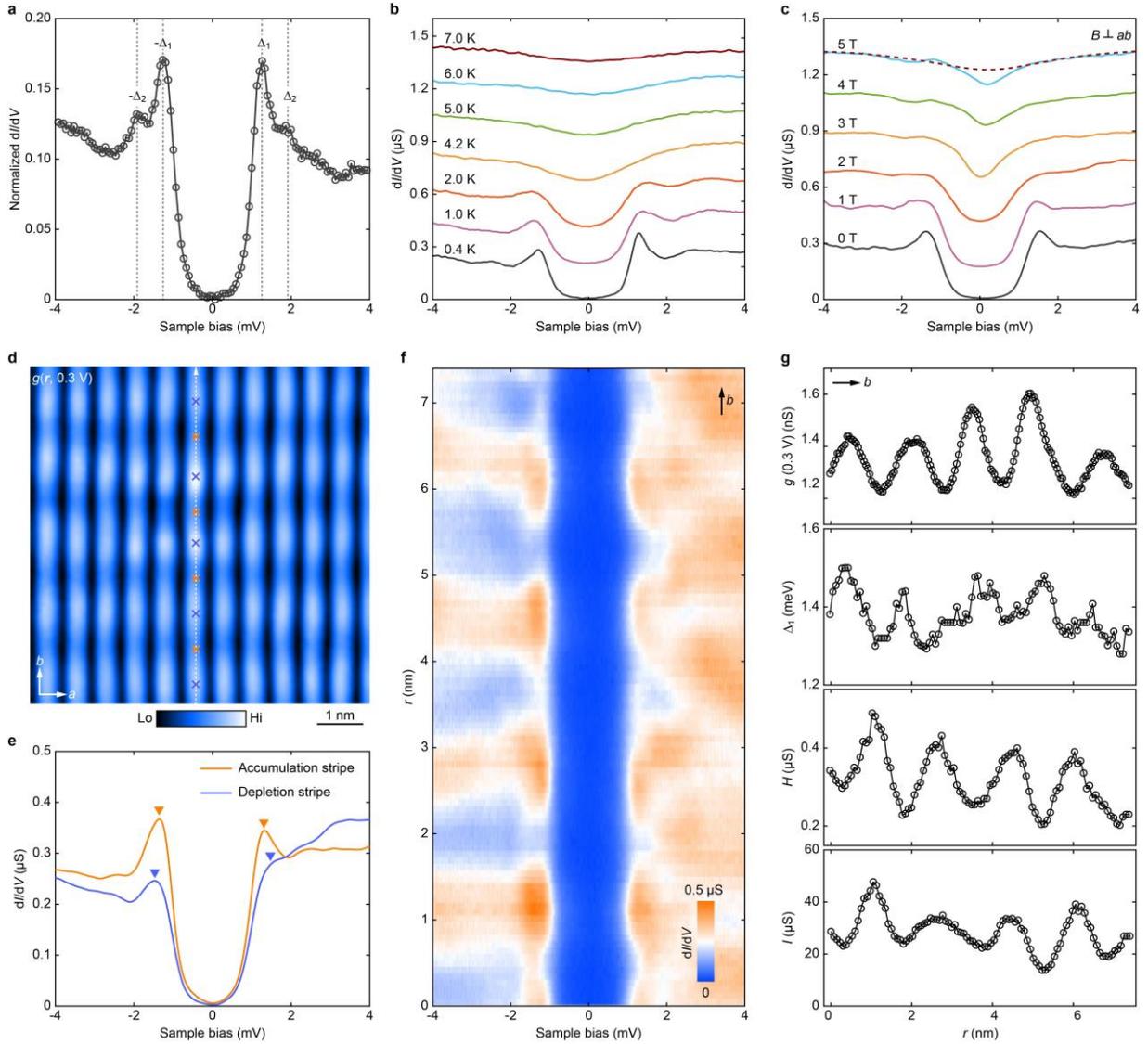

**Figure 3 | Striped PDW. a** Low-energy d$I$/d$V$ spectrum showing a two-gap superconductivity, defined as $\Delta_1$ and $\Delta_2$, in ML 1T′-MoTe$_2$. All superconducting gaps were acquired at 0.4 K, unless otherwise noticed, with a 4 MΩ junction resistance (setpoint: $V$ = 4 mV, $I$ = 1 nA). **b** Temperature-dependent d$I$/d$V$ spectra, revealing a $T_c$ ~ 6.0 K. **c** Site-specific d$I$/d$V$ spectra at various $B$ as indicated, showing a pseudogap-like feature above $B_{c2}$ ~ 4 T. The dashed curve represents the thermal broadening of the spectrum at 5 T (cyan) to 7.0 K. **d** Empty-state d$I$/d$V$ map $g(r, 0.3\,\text{eV})$ recorded on a 7.5 nm × 7.5 nm FOV, with alternating stripes of charge accumulation (orange crosses) and charge depletion (blue crosses). Setpoint: $V$ = 0.3 V, $I$ = 0.5 nA. **e** Two d$I$/d$V$ curves measured on the accumulation and depletion stripes, with the $\Delta_1$ gap edges marked by colored triangles. **f** Color map of the space-resolved d$I$/d$V$ spectra measured at equal separations (0.75 Å) along the white arrow in **d**. **g** Spatial modulations of $g(r, 0.3\,\text{eV})$, $\Delta_1(r)$, mean coherence peak height $H(r)$ and integrated spectral weight $I(r)$ within $\Delta_1(r)$.

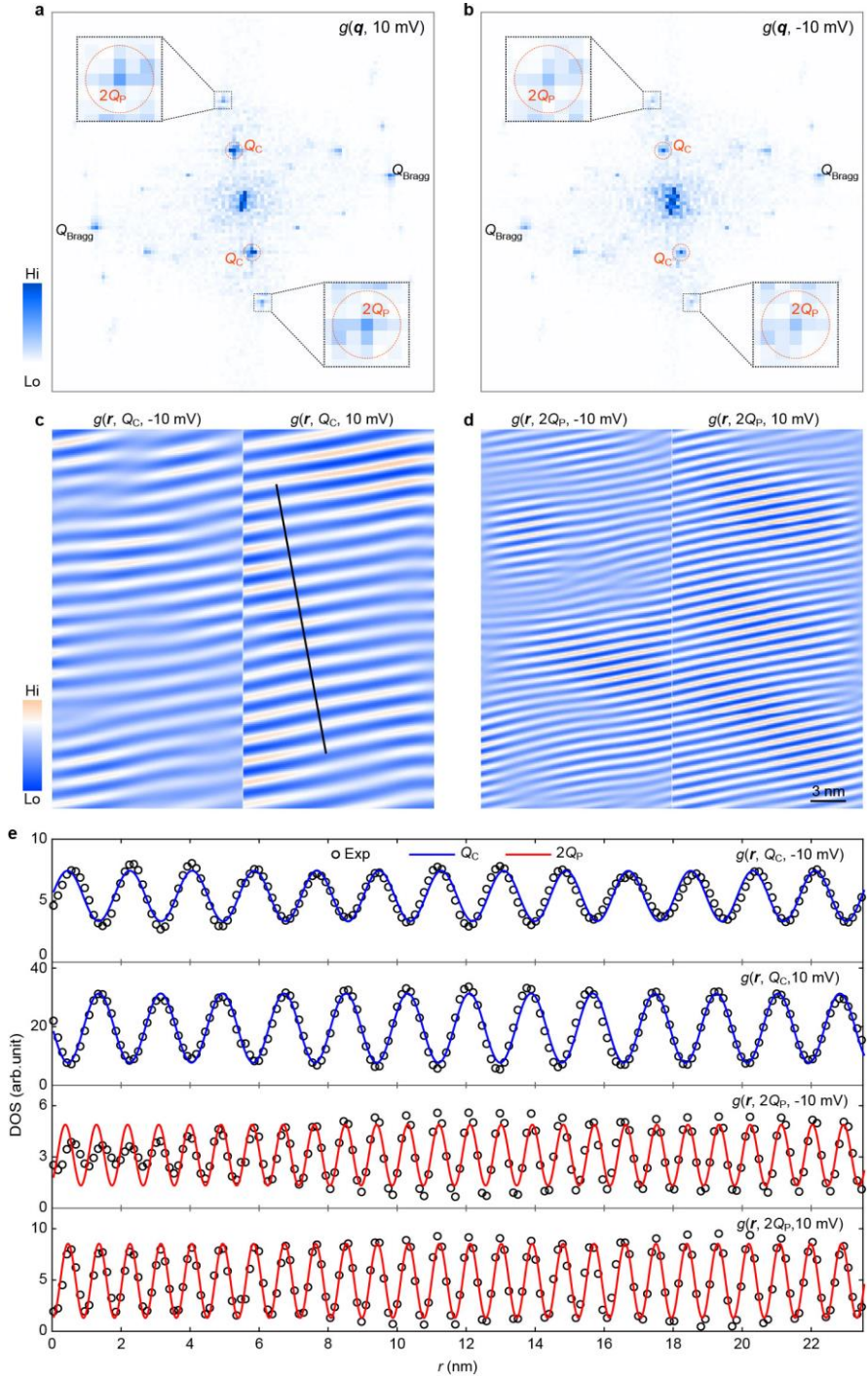

**Figure 4 | Particle-hole dichotomy between symmetry ($2Q_P$) and asymmetry ($Q_C$). a,b** Amplitudes of the Fourier transform of $g(r, \pm 10$ mV$)$, i.e., $g(q, \pm 10$ mV$)$. The outlined squares highlight the PDW-induced CO at $2Q_P$. **c,d** Gaussian-filtered $g(r, \pm 10$ mV$)$ maps in the filled (left) and empty (right) states by keeping wavevectors around $\pm Q_C$ and $\pm 2Q_P$ with a cutoff length of 0.06 Å$^{-1}$, marked by the red circles in **a,b**. **e** Profiles (black circles) of $g(r, Q_C, \pm 10$ mV$)$ and $g(r, 2Q_P, \pm 10$ mV$)$ measured along an identical trajectory marked by the black line in **c**. The blue and red lines denote the best fits of the experimental data to single cosineally modulated functions at $Q_C$ or $2Q_P$, respectively. In contrast to the intensity reversal between the $Q_C$-related $g(r, Q_C, \pm 10$ mV$)$, the PDW-induced CO at $2Q_P$ exhibits a unique particle-hole symmetry.

# Supplementary Information for

# Discovery of unidirectional charge and pair-density-wave orders

# in topological monolayer 1T′-MoTe$_2$

Li-Xuan Wei[1,†], Peng-Cheng Xiao[2,†], Fangsen Li[3,†], Li Wang[3], Bo-Yuan Deng[3], Fang-Jun Cheng[1], Fa-Wei Zheng[4], Ning Hao[5], Ping Zhang[2,6,*], Xu-Cun Ma[1,7,*], Qi-Kun Xue[1,7,8,9], Can-Li Song[1,7,*]

[1] *State Key Laboratory of Low-Dimensional Quantum Physics, Department of Physics, Tsinghua University, Beijing 100084, China*

[2] *School of Physics and Physical Engineering, Qufu Normal University, Qufu 273165, China*

[3] *Vacuum Interconnected Nanotech Workstation, Suzhou Institute of Nano-Tech and Nano-Bionics, Chinese Academy of Sciences, Suzhou 215123, China*

[4] *Key Lab of Advanced Optoelectronic Quantum Architecture and Measurement, School of Physics, Beijing Institute of Technology Beijing 100081, China*

[5] *Anhui Province Key Laboratory of Condensed Matter Physics at Extreme Conditions, High Magnetic Field Laboratory, HFIPS, Chinese Academy of Sciences, Hefei 230031, China*

[6] *Institute of Applied Physics and Computational Mathematics, Beijing 100088, China*

[7] *Frontier Science Center for Quantum Information, Beijing 100084, China*

[8] *Shenzhen Institute for Quantum Science and Engineering and Department of Physics, Southern University of Science and Technology, Shenzhen 518055, China*

[9] *Beijing Academy of Quantum Information Sciences, Beijing 100193, China*

[†]These authors contributed equally to this work.

*Correspondence to: clsong07@mail.tsinghua.edu.cn, zhang_ping@iapcm.ac.cn, xucunma@mail.tsinghua.edu.cn

**The PDF file includes:**
    Supplementary Notes (1-4)
    Figs. S1 to S19
    References

**Supplementary Notes**

1. Topological QSH edge states

A comparison of d$I$/d$V$($r$, $V$) spectra at many interiors and edges of ML 1T′-MoTe$_2$, as exemplified in Supplementary Fig. 2, reveals spectroscopic evidence of the edge states as prominent enhancements in the occupied states, albeit spatial inhomogeneity (Supplementary Fig. 2a,b). Plotted in Supplementary Fig. 2c are representative linecut d$I$/d$V$($r$, $V$) spectra from the step edge to interior. One can see that the edge states are predominantly located near 0.1 eV below $E_F$ and spread over a lateral dimension of ~ 2.8 nm. This can be further compellingly confirmed by acquiring the spectroscopic $g$($r$, $V$) maps across step edges of another ML 1T′-MoTe$_2$ island in Supplementary Fig. 3a. Plotted in Supplementary Fig. 3b are the spatially averaged horizontal profiles of the $T$($r$) (top row) and $g$($r$, $E = eV$) (bottom row) in the outlined area of Supplementary Fig. 3a. Apparently, the edge states develop in the vicinity of step edges and become pronounced below $E_F$, in good agreement with the $g$($r$, $V$) maps at various biases $V$ (Supplementary Fig. 3c-k). The unique states turn out to be robust against the varying edges of ML 1T′-MoTe$_2$ islands, irrespective of variations in edge geometry and chemical environment. The results imply the topologically nontrivial nature of the edge states protected by the time-reversal symmetry. This is further supported by computing the QSH states of the (100) edge in Supplementary Fig. 4, where the nontrivial topology of the edge states is corroborated by both spin polarization and a $Z_2$ topological number of 1 in Supplementary Fig. 5[1,2]. A careful inspection shows good agreement in energy between the QSH edge states in our experiments and DFT calculations. Note that the calculated QSH states have slight fluctuations in the energy distribution for various edge terminations. This offers a straightforward explanation of the inhomogeneous QSH edge states observed in Supplementary Fig. 2 and 3.

2. Superconductivity against interfacial disorder

As reported before[3], the thermally graphitized SiC(0001) surface is characterized by a large degree of disorder from the buried interfacial buffer layer. Such interfacial disorders are generally space-dependent and the strongest on ML graphene. In order to examine their potential impacts on the superconductivity, we have explored the heterointerface dependence of the superconducting gap $\Delta_1$ in ML 1T′-MoTe$_2$. As depicted in Supplementary Fig. 12a, the gap routinely becomes weakened with reducing graphene layer. In ML 1T′-MoTe$_2$ grown on the single-layer graphene with the strongest disorders, the superconducting energy gap $\Delta_1$ is hugely filled with ingap spectral weights, accompanied by significant suppression of the superconducting coherence peaks. Given that the work functions of graphene from 4.16 eV (monolayer) to 4.43 eV (trilayer) are close to that (~ 4.43 eV) of ML 1T′-MoTe$_2$[4], the interfacial charge transfer should be insignificant between the van der Waals stacked graphene and ML 1T′-MoTe$_2$. This excludes the heterointerface-dependent superconductivity by an interfacial electron transfer. Otherwise, one should have observed the optimal superconductivity in 1T′-MoTe$_2$ on single-layer graphene, because the largest electron transfer from the single-layer graphene benefits in superconductivity of ML 1T′-MoTe$_2$[5,6]. Alternatively, the interfacial disorders, which inevitably induce electron scattering that is harmful to superconductivity, provides a straightforward explanation for the significant suppression of $\Delta_1$ and its coherence peaks in ML 1T′-MoTe$_2$ grown on singe-layer graphene. For the heterostructures composed of ML 1T′-MoTe$_2$ and double-layer graphene (Supplementary Fig. 12b), the interfacial disorder effects are occasionally evident as well, giving rise to a wide distribution of $\Delta_1$ (Supplementary Fig. 12c). However, they occur on a length scale of a few tens of nanometers, significantly larger than that (~ 1.7 nm) of the PDW order.

3. Two distinct COs at $Q_C$ and $2Q_P$

In the Ginzburg-Landau theory[7,8], the coupled order parameter of a primary PDW at wavevector $Q_P$ with itself, i.e., $\Delta_Q \Delta^*_{-Q}$, can lead to a unique CO with twice of the PDW wavevector ($2Q_P$). In Fig. 2d, we find that the amplitude of Fourier transform of the $g(r, V)$ maps, i.e., $g(q, V)$, exhibits a distinct pair of peaks around $q \sim (0, \pm0.8)\pi/b \approx \pm 2Q_P$ for smaller $V$ from -20 mV to 70 mV, in addition to $Q_C \approx (0, 0.4)\pi/b$. They can be more clearly discernible in Supplementary Fig. 17. From the phase images of the Fourier transform of $g(r, V)$, we find a global initial phase difference $\delta\Phi = 2\Phi_P - \Phi_C = (0.58 \pm 0.2)\pi \sim 3\pi/5$ between the two distinct COs at $Q_C$ and $2Q_P$ (Supplementary Fig. 17c). This result, together with the observability of $2Q_P$ solely near $E_F$, reveals unambiguous evidences for a distinct CO for the $2Q_P$ peaks, other than an origin from the second harmonic of $Q_C$. Indeed, a closer inspection of the real-space $g(r, V)$ at smaller $V$ reveals complex spatial modulations beyond a single CO wavelength $\lambda \sim 5b$ (Fig. 2c), which are visualized by plotting the averaged d$I$/d$V(r, V)$ along the $b$ axis at four representative sample biases, i.e., $V = \pm 10$ mV and $V = \pm 20$ mV, in Supplementary Fig. 17d. Small peaks and/or humps-like features develop between the $Q_C$-driven DOS peaks, suggesting the involvement of another distinct CO modulation at the half wavelength. To describe this more quantitatively, we tentatively fit the spatial modulations of d$I$/d$V(r, V)$ along the crystal $b$ axis to a single cosineally-modulated function

$$dI/dV(r, V) \propto A_1 \cos(Q_C \cdot r + \Phi_C) + B \cdot r + C \qquad (2)$$

or a linear combination of two distinct cosineally-modulated functions

$$dI/dV(r, V) \propto A_1 \cos(Q_C \cdot r + \Phi_C) + A_2 \cos(2Q_P \cdot r + 2\Phi_P) + B \cdot r + C, \qquad (3)$$

where the term of $B \cdot r + C$ denotes the curved background, $\Phi_C$ and $2\Phi_P$ are initial phases of the two distinct COs, respectively. The results for the best fits are drawn as red lines in Supplementary Fig. 17d. Evidently, a single cosine function deviates from the spatial modulations of the experimentally measured d$I$/d$V(r, V)$, whereas the scenario of two coexisting COs describes the experimental data better, with a phase difference $\delta\Phi = 2\Phi_P - \Phi_C = 3\pi/5$ used.

4. Superconductivity at edges

Having established a unique coexistence of the superconductivity and QSH states in ML 1T′-MoTe$_2$, a more intriguing issue arises as to whether the superconductivity is retained at the step edges harboring the QSH states. Supplementary Fig. 19 illustrates one color map of the space-resolved d$I$/d$V$(***r***, $V$) spectra from the interior to step edge, taken along the *a*-axis direction to minimize the potential impact of the striped CO and PDW orders. Overlaid on the color map are two representative d$I$/d$V$ spectra measured near the edge and interior of ML 1T′-MoTe$_2$. It is interestingly found that the superconducting gap continuously extends to the step edges and gets only weakly suppressed in the gap magnitude $\Delta_1$ and coherence peaks. The results indicate that we have realized a potential connate 1D topological superconductor along the step edges of the 1T′-MoTe$_2$ films. This merits further investigations.

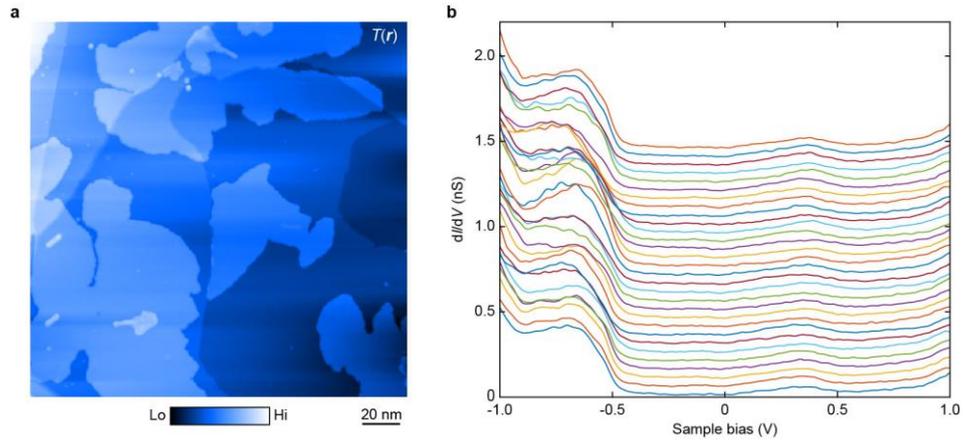

**Figure S1. STM characterization of 1T′-MoTe$_2$ epitaxial films. a** Representative large-scale STM morphology $T(r)$ (200 nm × 200 nm, $V$ = 2.0 V, $I$ = 10 pA) of 1T′-MoTe$_2$ epitaxial films prepared on graphitized 4$H$-SiC(0001) substrates. Monolayer 1T′-MoTe$_2$ islands with lateral dimension of several tens of nanometers can be readily obtained and are more or less decorated by tiny sheets of bilayer MoTe$_2$. **b** Uniformity of the tunneling d$I$/d$V$ spectra on a large energy scale of ±1.0 eV, measured at equal separations (~ 0.47 nm) along a trajectory of 14 nm on ML 1T′-MoTe$_2$. Set point: $V$ = 1.0 V, $I$ = 0.5 nA.

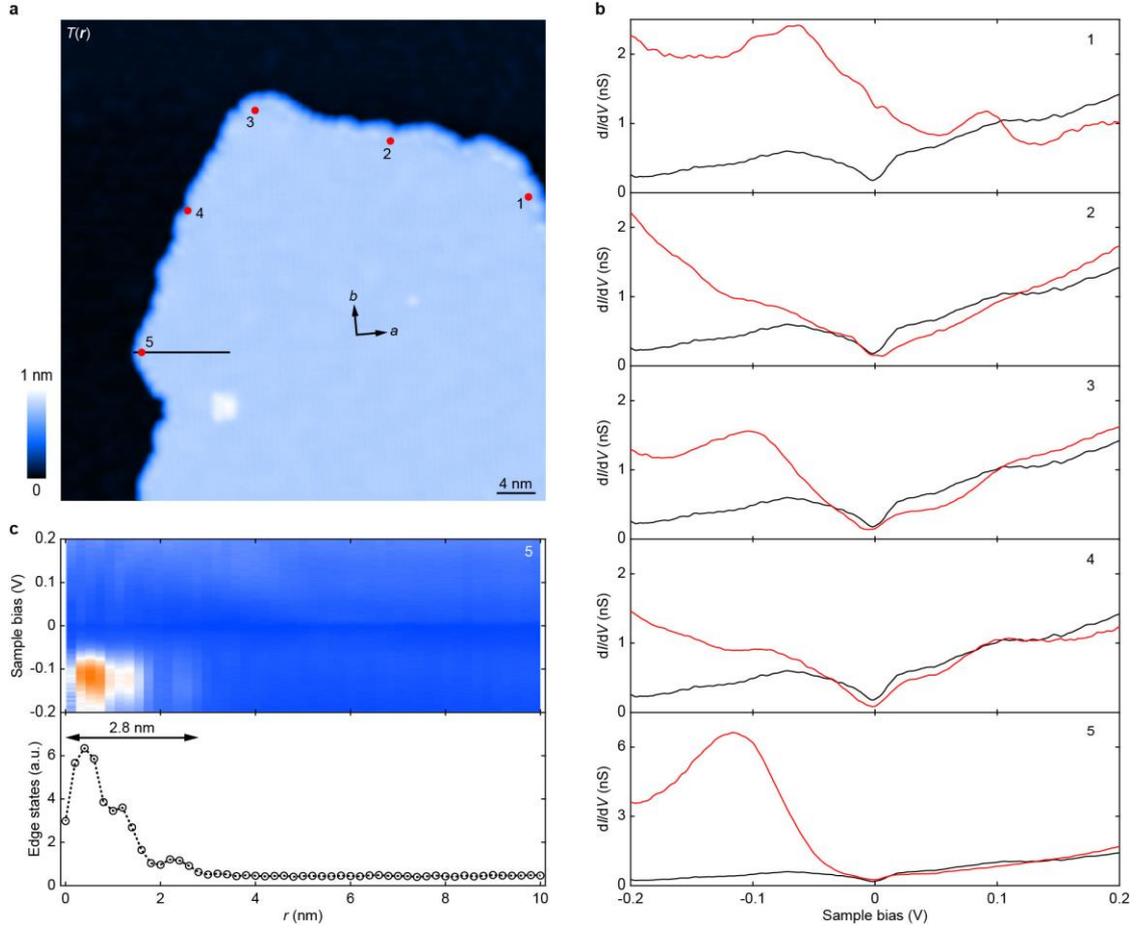

**Figure S2. Topological QSH states at edges. a** STM topography $T(r)$ (50 nm × 50 nm, $V$ = 2.0 V, $I$ = 10 pA) showing one irregular step edge of ML 1T′-MoTe$_2$. **b** Tunneling d$I$/d$V$ spectra (red curves) measured at five different locations (red dots in **a**) at edges. For comparison, one d$I$/d$V$ spectroscopy at the interior of ML 1T′-MoTe$_2$ is plotted as black curves. Set point: $V$ = 0.2 V, $I$ = 0.2 nA. **c** Color map (top row) of the linecut d$I$/d$V(r, V)$ spectra acquired at equal separations (0.2 nm) along a trajectory from the step edge to the island interior marked by the line in **a**. The bottom row quantifies the space $r$-dependent edge states by the integrated conductance $g(r, -150$ mV $< V < -50$ mV$)$. The edge states exist locally at step edges on a length of 2.8 nm.

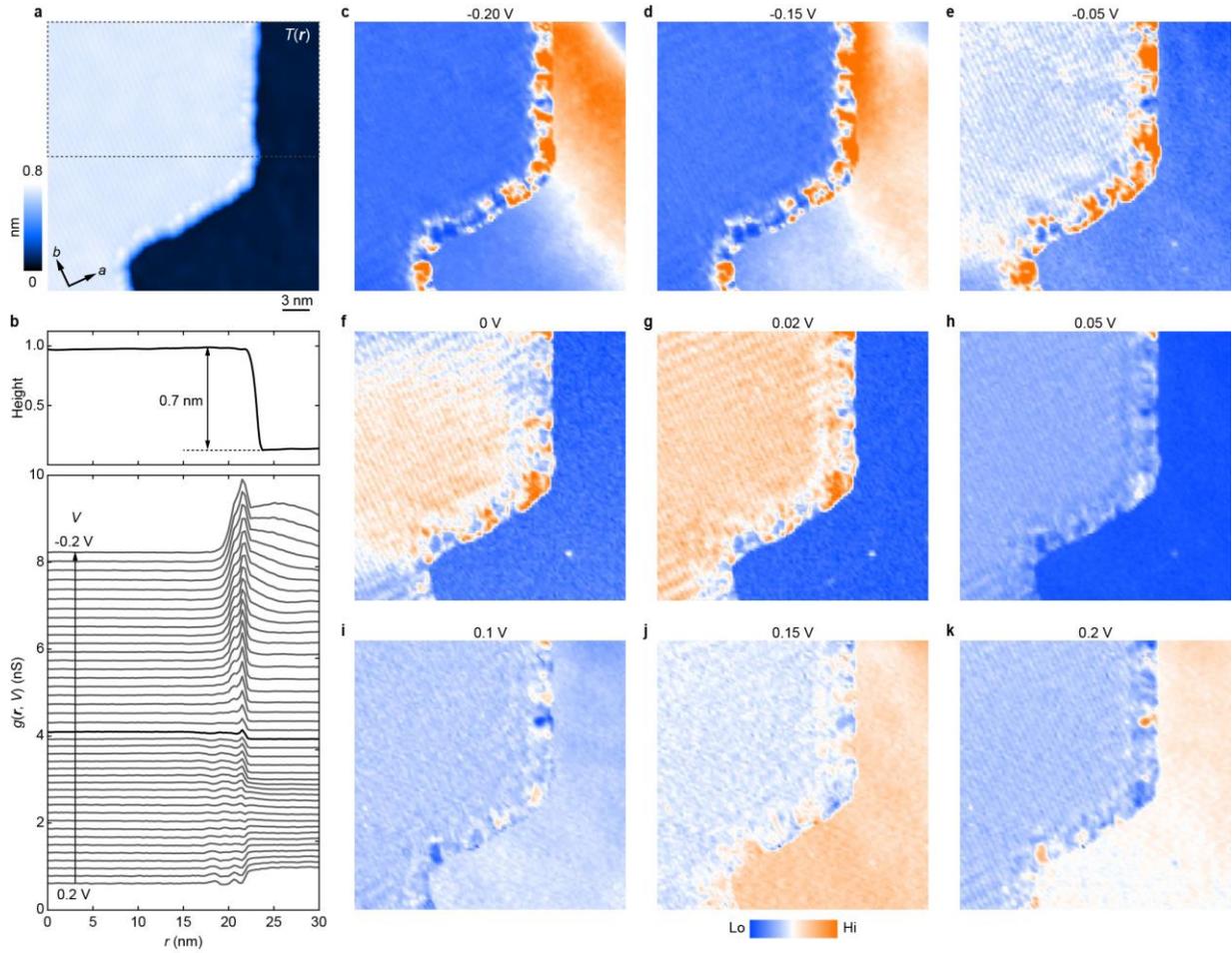

**Figure S3. Mapping the QSH states. a** Another $T(r)$ (30 nm × 30 nm, $V = 1.0$ V, $I = 10$ pA) presenting a relatively regular step edge of ML 1T′-MoTe$_2$. **b** Means of the horizontal profiles of $T(r)$ (top row) and energy-dependent $g(r, E = eV)$ (bottom row) in the outlined rectangle of **a**. The thick black line highlights the $g(r, V)$ profile at $E_F$ ($V = 0$ mV), below which the edge states start to increase in a prominent manner. **c-k** Simultaneously measured $g(r, E = eV)$ maps at various biases $V$ as indicated, in the same FOV as **a**. Set point: $V = 0.2$ V, $I = 0.1$ nA.

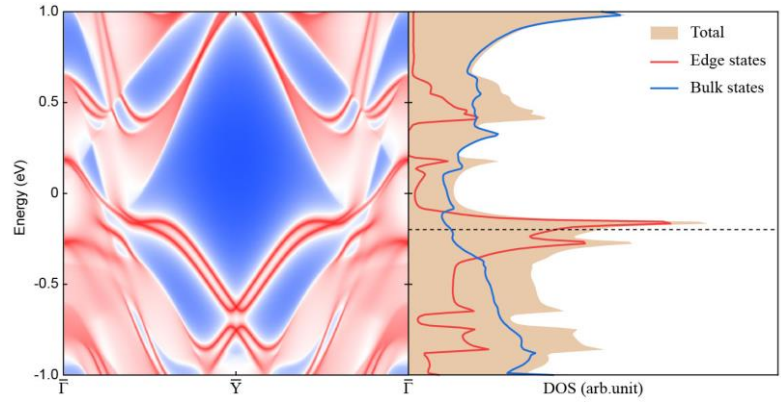

**Figure S4. Calculated QSH states at the (100) edges.** The QSH states reside inside the bulk gap of ML 1T′-MoTe$_2$ at the $\overline{Y}$ point (left), with the integrated DOS from the bulk states (blue) and (100) edge states (red) shown in the right panel. The orange-shaded region denotes the total DOS.

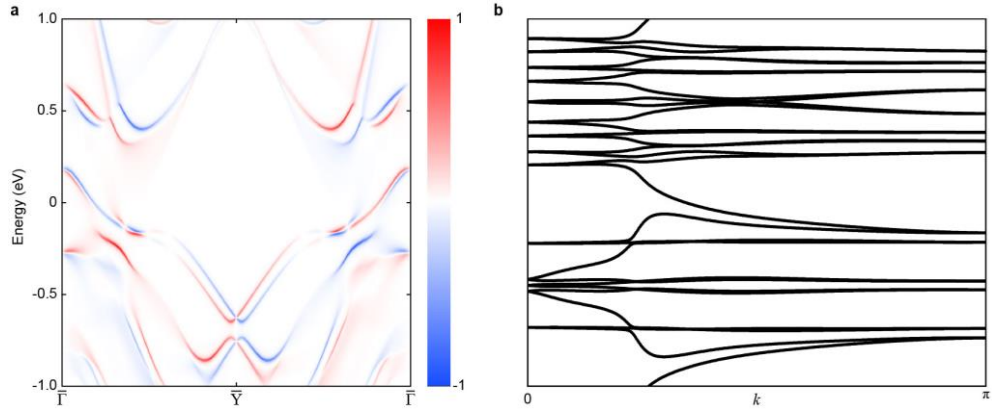

**Figure S5. Nontrivial topological edge states. a** Spin polarization of the topological edge states in ML 1T′-MoTe$_2$. **b** Wannier charge centers (Wilson loop) of ML 1T′-MoTe$_2$ for the time-reversal invariant momentum plane $k_z = 0$. The Z$_2$ topological number of 1 indicates the nontrivial topological nature of the edge states.

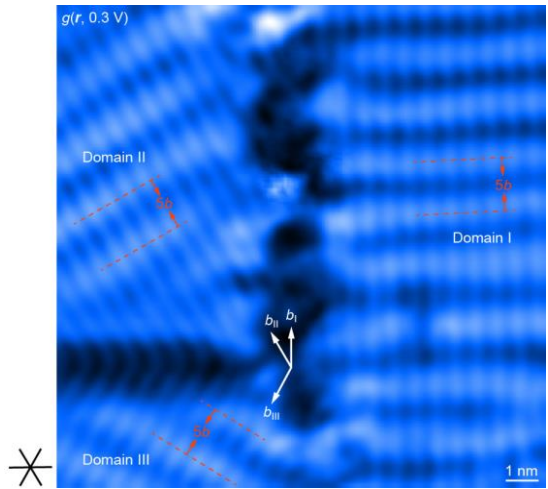

**Figure S6. CO against twist angles between the 1T′-MoTe$_2$ and graphene.** Space-resolved conductance map d$I$/d$V$($r$, $V$) acquired at $V$ = 0.3 V in a 15 nm × 15 nm FOV, with three distinct domains of ML 1T′-MoTe$_2$ marked by I, II and III, respectively. The black crossing lines denote three equivalent orientations of the carbon-carbon bonds (i.e., $Q_{6\times6}$) of buried graphene, rotated by 30º relative to the graphene lattice. Guided by the zig-zag lattice structures along the crystal $b$ axis ($b_\mathrm{I}$, $b_\mathrm{II}$ and $b_\mathrm{III}$) of 1T′-MoTe$_2$, we can determine the twist angles between the underlying 6 × 6 superstructure and ML 1T′-MoTe$_2$ lattices in domain I (15º), domain II (1º) and domain III (2º). Despite this apparent variation, the CO is invariably oriented along the $b$ axis and shows a consistent periodicity of ~ 5$b$, as marked by the parallel dashes. Set point: $V$ = 0.3 V, $I$ = 1.0 nA.

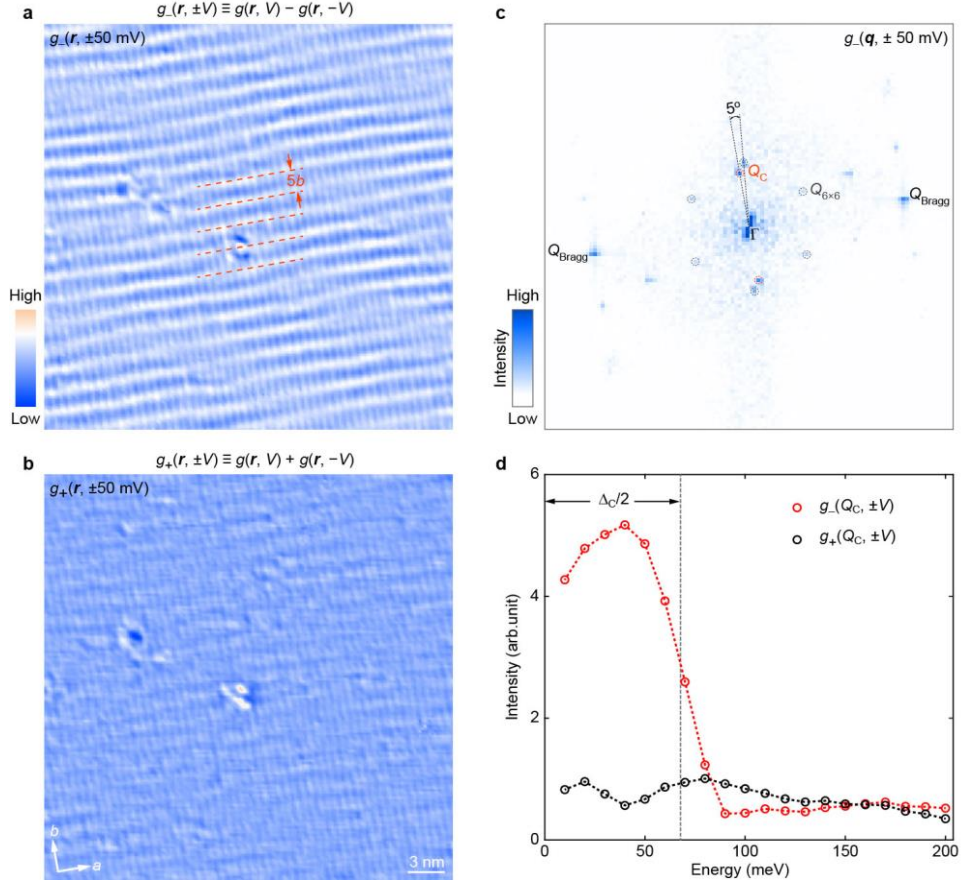

**Figure S7. CO constant inversion between opposite biases. a,b** Spectroscopic imaging (33 nm × 33 nm) of the digital subtraction (**a**) and addition (**b**) between the simultaneously measured empty-state $g(r, 50\ \text{mV})$ and filled-state $g(r, -50\ \text{mV})$ maps, defined as $g_-(r, \pm 50\ \text{mV}) \equiv g(r, 50\ \text{mV}) - g(r, -50\ \text{mV})$ and $g_+(r, \pm 50\ \text{mV}) \equiv g(r, 50\ \text{mV}) + g(r, -50\ \text{mV})$, respectively. The spatial modulation of CO is distinctly enhanced in **a**, but becomes barely visible in **b**. This indicates a marked contrast inversion of the CO between opposite biases. Set point: $V = 0.3$ V, $I = 1.0$ nA. **c** Amplitude of the Fourier transform of $g_-(r, \pm 50\ \text{mV})$, i.e., $g_-(q, \pm 50\ \text{mV})$, exhibiting prominent CO peaks at $Q_C \approx (0, 0.4)\pi/b$. The twist angle is estimated to be about 5° between the underlying 6 × 6 and 1T′-MoTe$_2$ lattices. **d** CO peak intensities $g_-(Q_C, V)$ (red) and $g_+(Q_C, V)$ (black) as a function of the energy $E = eV$, calculated from the amplitudes $g_-(q, V)$ and $g_+(q, V)$ of Fourier transforms of the $g_-(r, V)$ and $g_+(r, V)$, with the sample bias $V$ changing from 10 mV to 200 mV. The vertical dashes correspond to the edge energy of CO-opened gap, i.e., $\Delta_C/2 \sim 70$ meV, below which the CO contrast inversion becomes prominent.

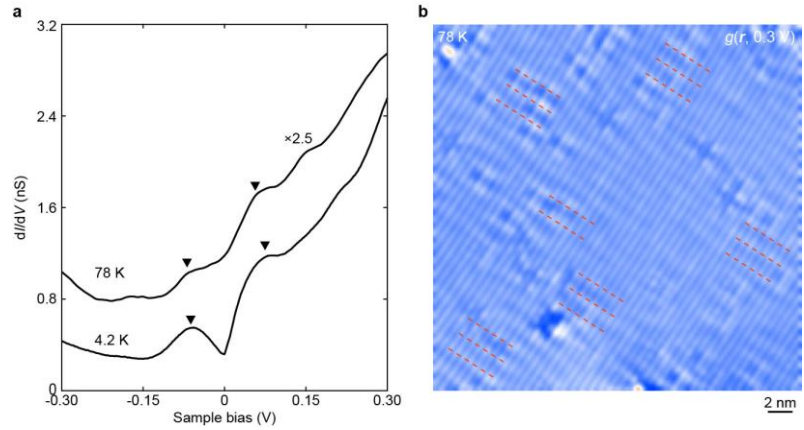

**Figure S8. Unidirectional CO at elevated temperature. a** Spatially-averaged d$I$/d$V$ spectra presenting energy gap opening near $E_F$ by the CO at 4.2 K (Set point: $V = 0.2$ V, $I = 0.5$ nA) and 78 K (Set point: $V = 0.3$ V, $I = 0.2$ nA). For clarity, the spectrum taken at 78 K has been multiplied by 2.5 and offset vertically by 0.5 nS. The triangles mark edges of the CO-opened gaps $\Delta_C$. **b** $g(r, V = 0.3$ V$)$ map (30 nm × 30 nm) measured at 78 K. It shows a substantial melting of the long-range CO into short-range striped patterns, as marked by the red parallel dashes. Setpoint: $V = 0.3$ V, $I = 0.5$ nA.

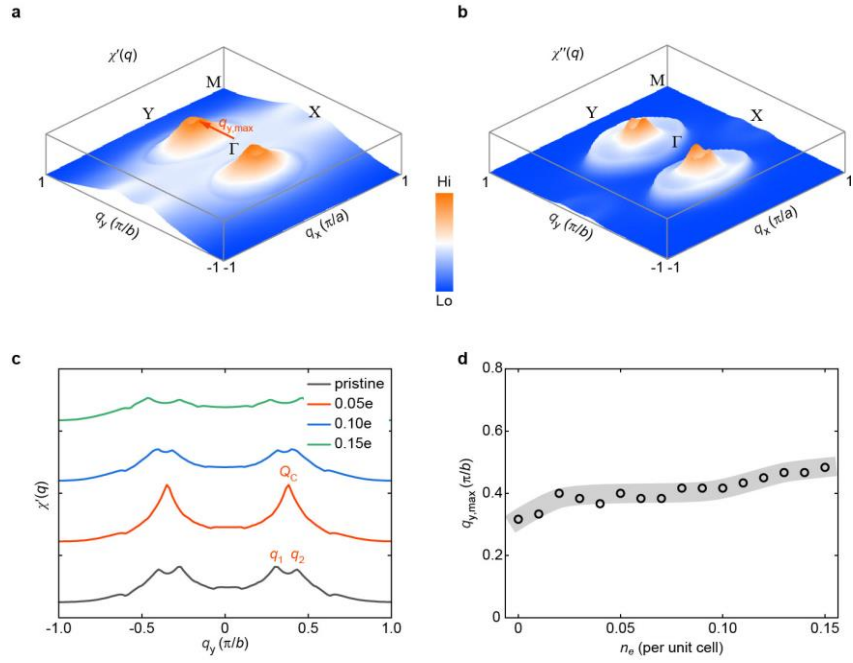

**Figure S9. $\chi(q)$ and its leading eigenvalues with electron doping. a,b** Real ($\chi'(q)$) and imaginary ($\chi''(q)$) parts of electron susceptibility between the electron and hole pockets crossing the $E_F$ in pristine ML 1T′-MoTe$_2$, respectively. The color bar marks the relative value. The leading eigenvalue of $\chi'(q)$ is indicated by an arrow at a wavevector $q_{y,max}$. **c** Electron doping-dependent linecut of $\chi'(q)$ along the $q_y$ direction with $q_x = 0$. Peaks correspond to the inter-band nesting vectors ($q_1$ and $q_2$) between the electron and hole pockets on the Fermi surface of ML 1T′-MoTe$_2$. **d** The evolution of $q_{y,max}$ *versus* the electron doping level $n_e$. Note that a leading eigenvalue of $q_{y,max} \sim (0, 0.4)\pi/b$ occurs robustly at a broad electron doping from $n_e \sim 0.02$ to $n_e \sim 0.10$. The shaded ribbon is guide to the eye.

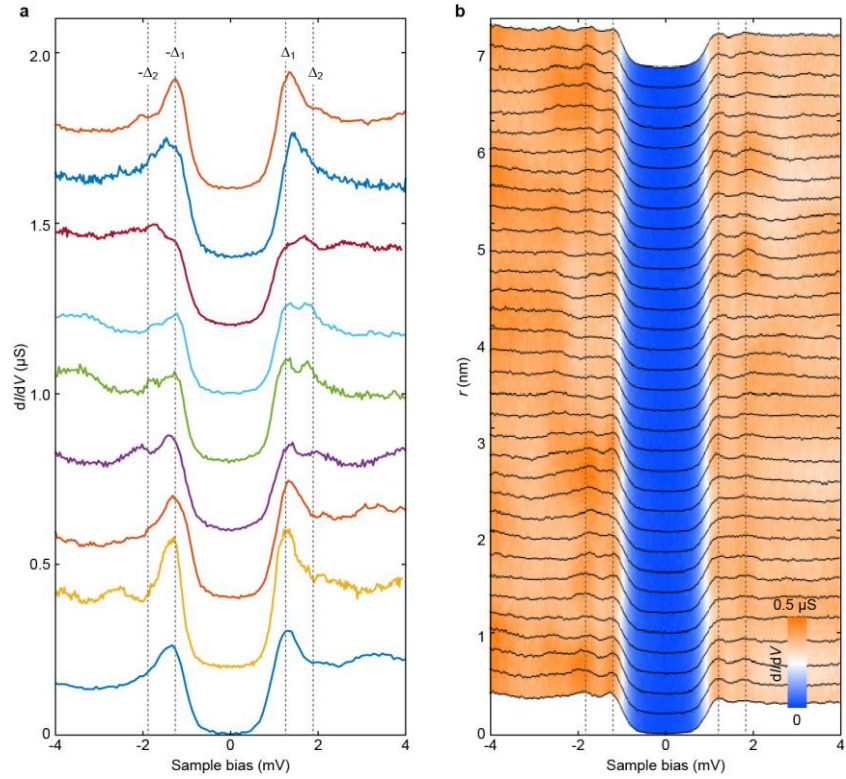

**Figure S10. Two-gap superconductivity. a** Representative d$I$/d$V$ spectra measured at different locations in ML 1T′-MoTe$_2$. Although both superconducting coherence peaks corresponding to $\Delta_1$ are robustly visible, the gap edges of the larger $\Delta_2$ are not consistently discernible. This likely originates from a relatively weaker contribution of $\Delta_2$ than $\Delta_1$ due to a band-selective tunneling matrix element effect and/or the proximity between $\Delta_2$ and $\Delta_1$. The spectra have been vertically offset for clarity. **b** Color map of the linecut d$I$/d$V$ spectra on region clearly exhibiting two superconducting gaps. Vertical dashes mark the magnitudes of the superconducting gaps $\Delta_1$ and $\Delta_2$. Additional peak (hump) structures can be seen outside the $\Delta_2$ edges and are mostly asymmetric with $E_F$. Their origin remains to be unknown at present.

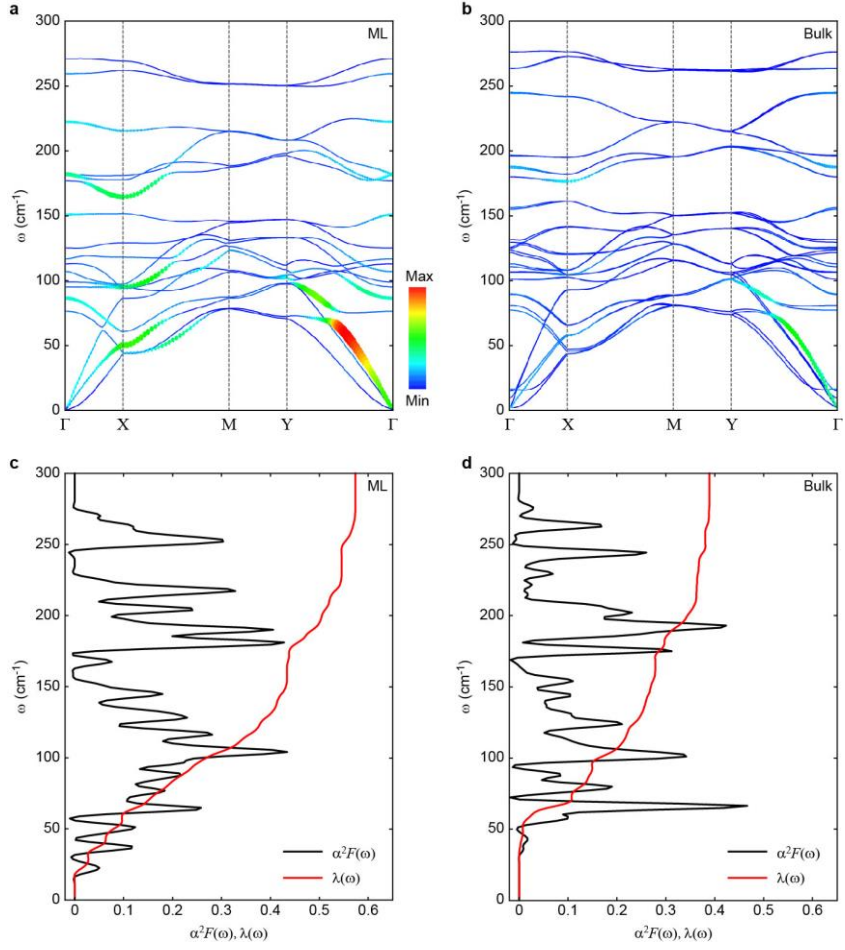

**Figure S11. Electron-phonon coupling in ML 1T′-MoTe$_2$. a,b** Phonon spectra $w_{qv}$ with the wave vector $q$ and mode index $v$ in (**a**) ML 1T′-MoTe$_2$ and (**b**) its bulk counterpart T$_d$-MoTe$_2$. Albeit without phonon softening, certain phonon spectra exhibit significantly enlarged linewidth $\lambda_{qv}$, marked by the weight of the curves, as the dimensionality is reduced from the bulk to the ML limit. **c,d** Eliashberg spectral function $\alpha^2F(w)$ and electron-phonon coupling strength $\lambda(w)$ in ML 1T′-MoTe$_2$ and its bulk counterpart, respectively. As the dimensionality reduces, the $\lambda(w)$ is evidently enhanced, consistent with the observed $T_c$ enhancement in ML 1T′-MoTe$_2$.

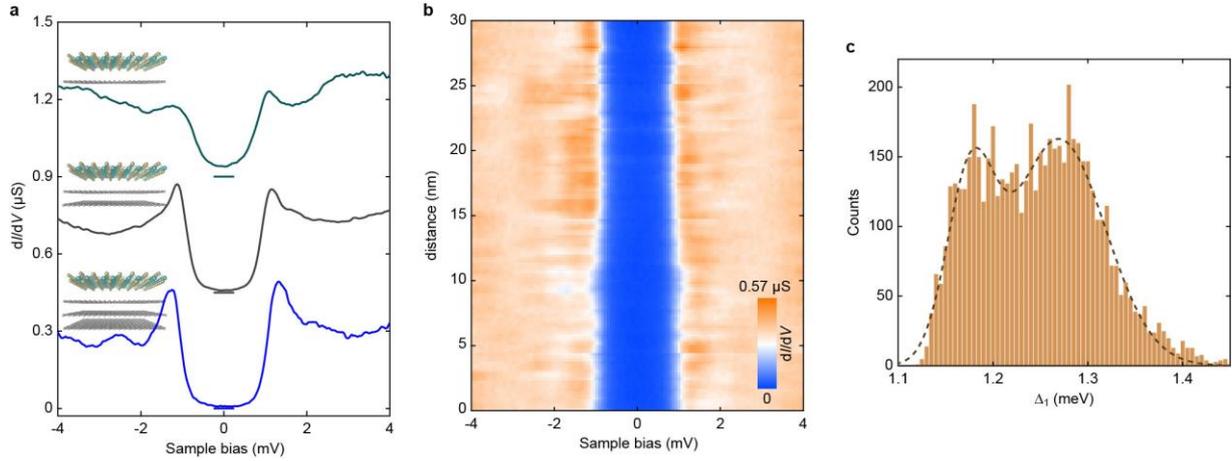

**Figure S12. Superconductivity against interfacial disorder. a** Representative d$I$/d$V$ spectra of ML 1T′-MoTe$_2$ grown on single-layer (green), bilayer (grey) and trilayer (blue) graphene substrates as sketched in insets. The gray and green spectra have been vertically offset for clarity, with the zero conductance positions marked by the correspondingly colored horizontal bars. The superconductivity becomes so weakened with reducing graphene layer that a V-shaped gap with significantly filled subgap DOS occurs in ML 1T′-MoTe$_2$ grown on the roughest single-layer graphene with significant buried disorder. **b** Linecut d$I$/d$V$ spectra taken at equal separations (0.3 nm) along a trajectory of 30 nm in ML 1T′-MoTe$_2$ grown on the bilayer graphene. The spatial inhomogeneity is evident and caused primarily by space-dependent disorders from the substrate. **c** Histogram of space-resolved $\Delta_1(\boldsymbol{r})$, half the distance between the two coherence peaks, deduced from 5760 d$I$/d$V$ spectra in ML 1T′-MoTe$_2$ grown on bilayer graphene. The $\Delta_1(\boldsymbol{r})$ distribution can be described by a double-Gaussian function peaked at 1.18 meV and 1.27 meV, shown as the dashed curve.
1616

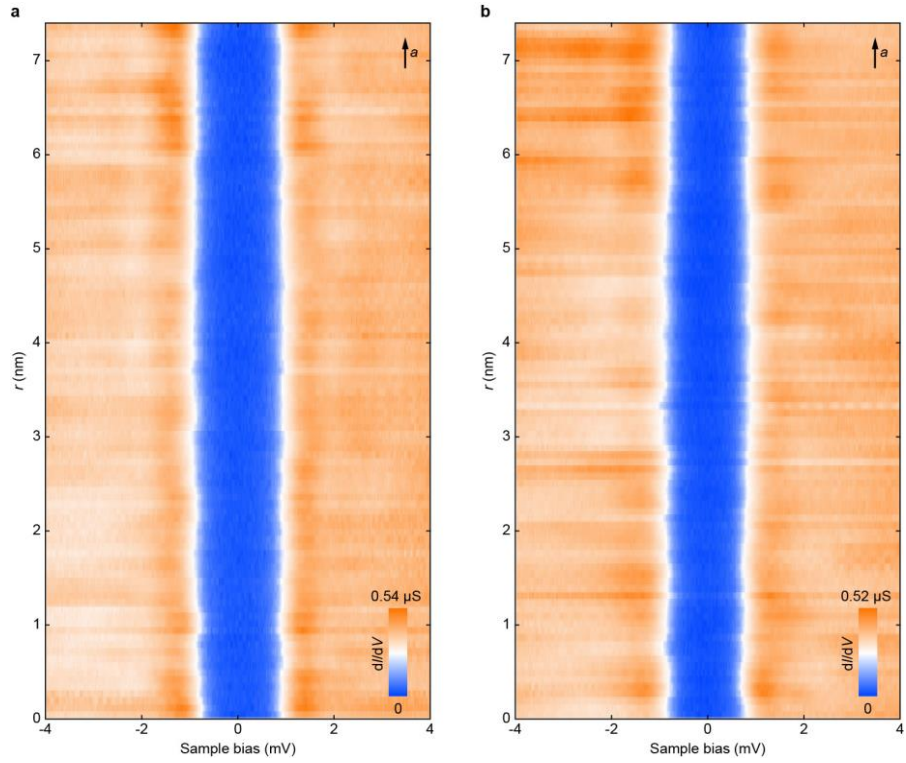

**Figure S13. Absence of periodic modulation in $\Delta_1(r)$ along the crystal *a* axis. a,b** Color maps of line-cut d$I$/d$V$ spectra measured at equal separations (0.75 Å) along the charge accumulation and charge depletion stripes, respectively. No periodic modulation is visible in $\Delta_1(r)$, as opposed to those along the crystal *b* axis.

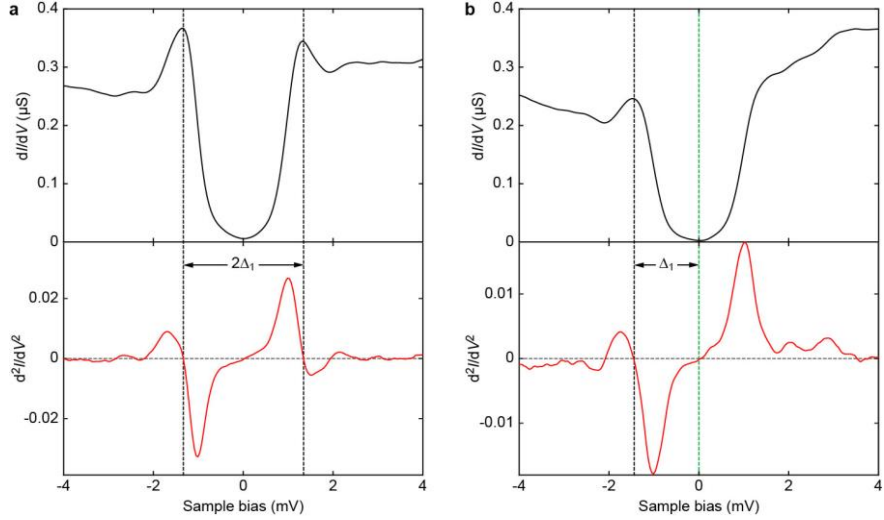

**Figure S14. Extraction of the superconducting gap magnitude $\Delta_1$. a** Representative d$I$/d$V$ spectrum (top panel) with two apparent coherence peaks and its derivative d$^2I$/d$V^2$ (bottom panel) as a function of energy. The black vertical dashes mark the energy positions of the two coherence peaks, at which the d$^2I$/d$V^2$ curves reach zero. This allows us to readily extract $\Delta_1$ as half-distance between the two coherence peaks. **b** Typical d$I$/d$V$ spectrum (top panel) with only one visible coherence peak and its derivative d$^2I$/d$V^2$ (bottom panel) as a function of energy. Given that it is not straightforward to define the superconducting gap from the empty state, we extract $\Delta_1$ as the energy of the discernible coherence peak with respect to $E_F$ (marked by the green dashed line), thanks to the particle-hole symmetry in the superconducting gap.

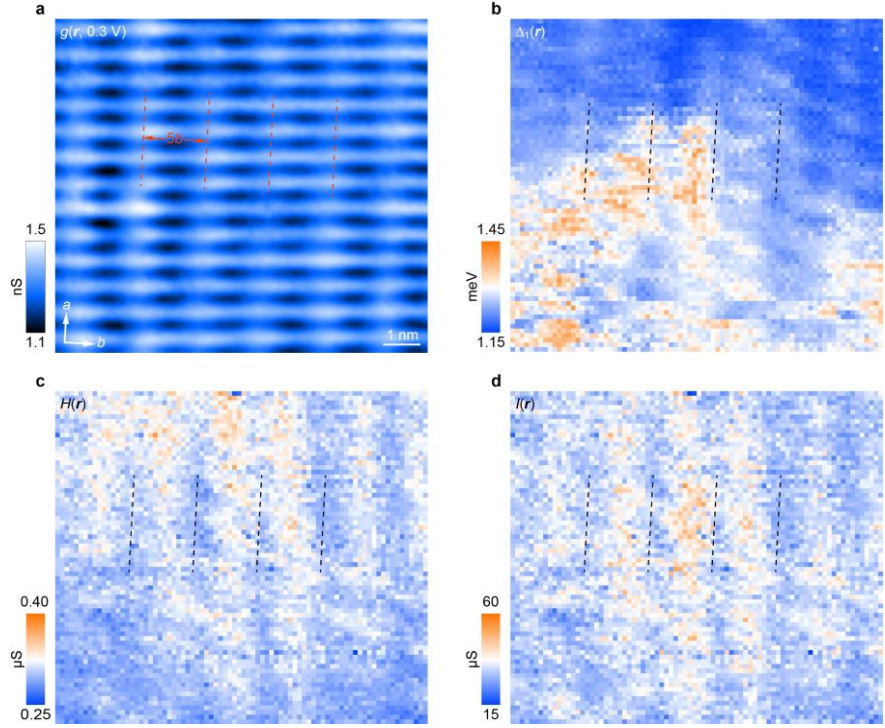

**Figure S15. Mapping the striped PDW order. a** Empty-state $g(r, V = 0.3$ eV$)$ taken on a 10.0 nm × 9.0 nm FOV, over which a large set of d$I$/d$V$ spectra have been measured on equidistant grid sites (80 pixels × 72 pixels) to image the spatial modulation of $\Delta_1(r)$ in real space. Set point: $V = 0.3$ V, $I = 1.0$ nA. **b-d** Maps of the superconductivity parameters $\Delta_1(r)$, $H(r)$ and $I(r)$ extracted from the grid d$I$/d$V(r, V)$ spectra, respectively. The red and black dashes denote the stripes of charge depletion and $\Delta_1$ maxima, respectively.

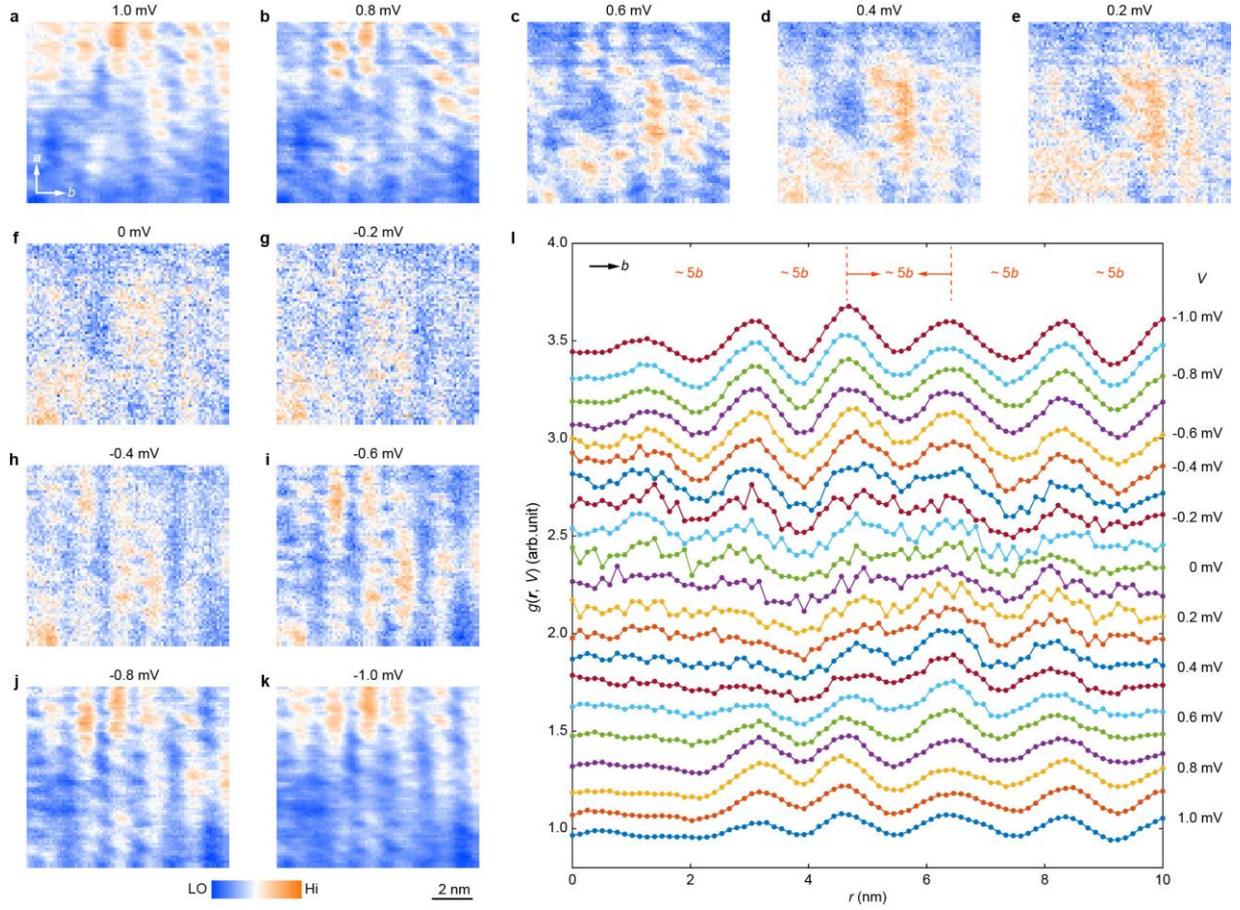

**Figure S16. Spatial modulations of the raw data $g(r, |V| \leq 1$ mV). a-k** Energy-dependent $dI/dV(r, E = eV)$ maps, measured in the same FOV as Supplementary Fig. 11a, with $|V| \leq 1$ mV (as indicated) locating within the superconducting energy gap $\Delta_1$. **l** Averaged line profiles of $dI/dV(r, V)$ along the crystal $b$ axis at various sample biases $V$. The spatial modulations of $g(r, V)$, namely the striped PDW order, can be seen even for $V = 0$ mV. In contrast to the CO, there exists no contrast inversion of the PDW state between opposite biases, a hallmark of particle-hole symmetry.

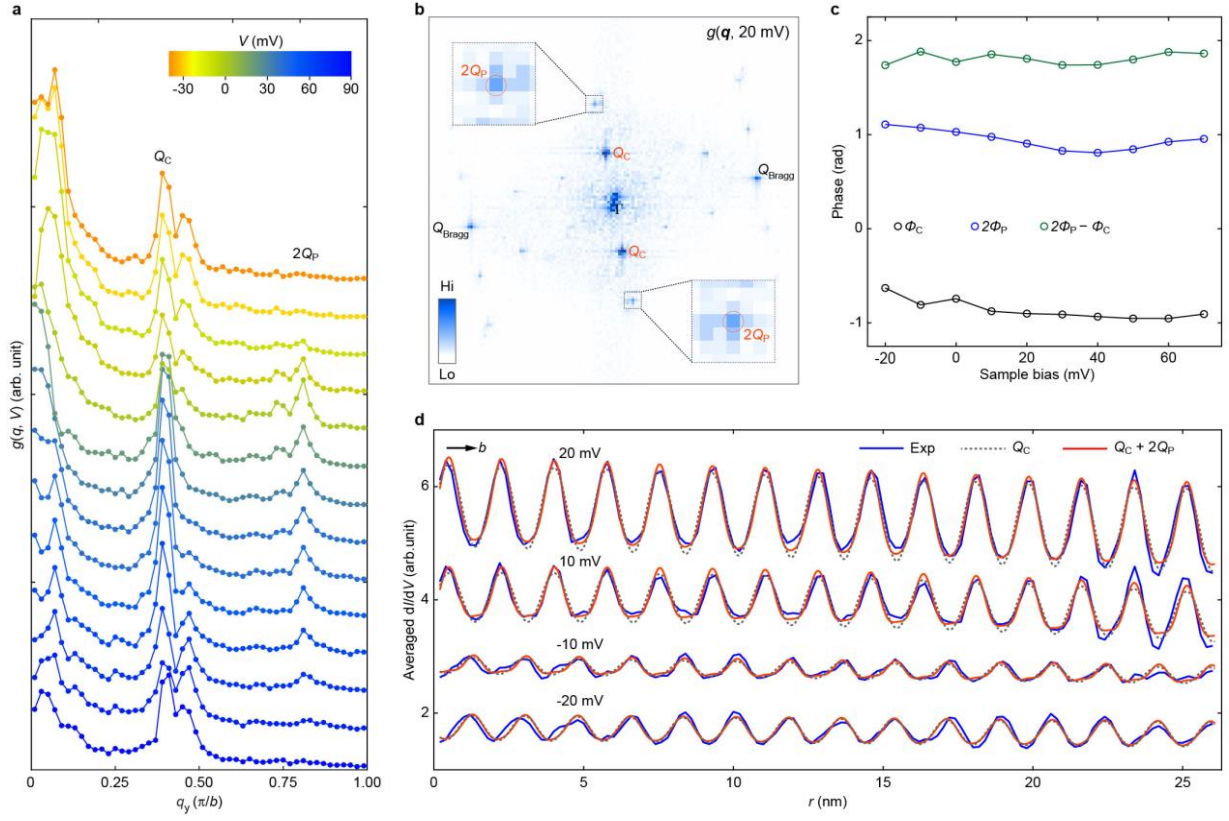

**Figure S17. Dichotomy between the COs at $Q_C$ and $2Q_P$. a** Amplitude profiles of the Fourier transform of $g(\mathbf{r}, V)$ for smaller $V$, measured along the $q_y$ direction. In addition to $Q_C \approx (0, 0.4)\pi/b$, a distinct pair of peaks develop at $2Q_P \approx (0, 0.8)\pi/b$ for -20 meV $\leq E \leq$ 70 meV. **b** Amplitude of the Fourier transform $g(\mathbf{r}, 20$ mV), i.e., $g(\mathbf{q}, 20$ mV), showing two wavevectors $Q_C$ for the 5$b$ CO and $2Q_P$ induced by the 5$b$ PDW order. Insets show zoom-in of the outlined areas, more clearly displaying the $2Q_P$ wavevector. **c** Initial phases of $\Phi_C$ (black circles) and $2\Phi_P$ (blue circles) obtained from the phase of Fourier transform of $g(\mathbf{r}, V)$ at various $V$, as well as the phase difference $\delta\Phi' = 2\Phi_P - \Phi_C$ (green circles). Both $\Phi_C$ and $2\Phi_P$ display small variations with the sample bias $V$ that get significantly weaker for $\delta\Phi'$. **d** Averaged $g(\mathbf{r}, V)$ profiles along the crystal $b$ axis (blue lines) at various $V$ as indicated. For clarity, the top three curves are vertically shifted by 1, 2 and 3 from bottom to top. Gray dotted lines and red lines designate the best fits of the experimental data to a single cosineally modulated function and a linear combination of two distinct cosineally modulated functions, respectively.

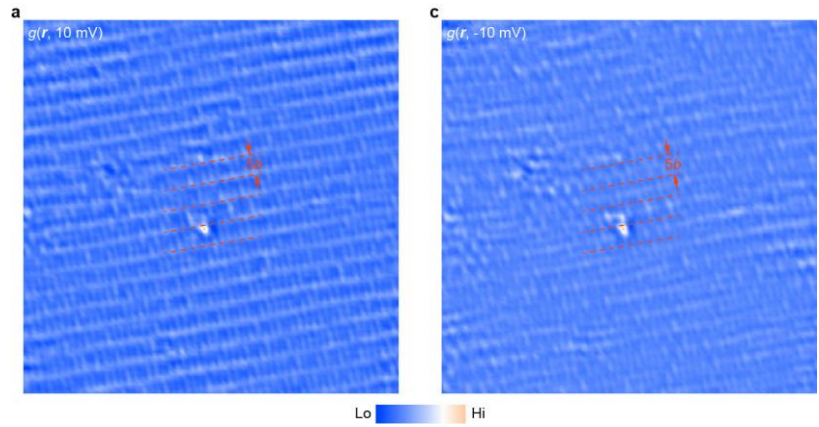

**Figure S18. Raw conductance maps. a** Empty-state $g(\mathbf{r}, 10\text{ mV})$ map in the same FOV of 33 nm × 33 nm as Fig. 2c. **b** Same to **a**, but for the filled-state at $V = -10$ mV.

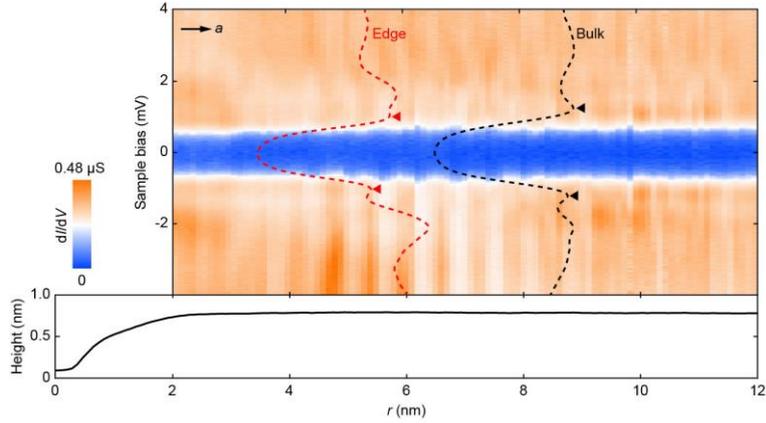

**Figure S19**. **Edge superconductivity**. Color map of the space-resolved superconducting energy gaps (top row) along a 10 nm trajectory from one step edge to the interior as shown in the $T(r)$ profile (bottom row). Overlaid are two representative d$I$/d$V$ spectra taken at the edge (red) and interior (black) of ML 1T′-MoTe$_2$. The colored triangles mark the gap edges of $\Delta_1$.